\documentclass[twocolumn]{aastex631}
\usepackage{tabularx}
\usepackage{makecell}
\usepackage{graphicx}

\usepackage{hyperref}
\usepackage{array}
\usepackage{makecell}
\usepackage{graphicx}

\def\kms{\ensuremath{\rm{\,km\,s^{-1}}}}
\def\fesc{$f_{\rm{esc}}$}
\def\lya{Ly$\alpha$}
\def\fcov{$f_{\rm{cov}}$}
\def\fcovlis{$f_{\rm{cov}}(\rm{LIS})$}
\def\fcovhi{$f_{\rm{cov}}(\rm{HI})$}

\newcommand{\Lya}{\ensuremath{\rm Ly\alpha}}

\newcommand{\Hb}{\ensuremath{\rm H\beta}}
\newcommand{\Oiii}[1]{[\ion{O}{3}] \ensuremath{#1}}

\newcommand{\Civ}{\ion{C}{4}}

\newcommand{\Hi}{\ion{H}{1}}

\newcommand{\Siiiabs}[1]{\ion{Si}{2} \ensuremath{#1}}
\newcommand{\Ciiabs}[1]{\ion{C}{2} \ensuremath{#1}}
\newcommand{\Oiabs}[1]{\ion{O}{1} \ensuremath{#1}}

\begin{document}

\title{Stacking PANCAKEZ: sPectroscopic Analysis with NirspeC stAcKs in the Epoch of reioniZation. Weak ISM Absorption and Implications for Ionizing Photon Escape at $z\sim7$}

\author[0000-0002-4453-5870]{Kelsey S. Glazer}
\affiliation{Department of Physics and Astronomy, University of California, Davis, 1 Shields Ave, Davis, CA 95616, USA}

\author[0000-0001-5860-3419]{Tucker Jones}
\affiliation{Department of Physics and Astronomy, University of California, Davis, 1 Shields Ave, Davis, CA 95616, USA}

\author[0000-0003-4520-5395]{Yuguang Chen}
\affiliation{Department of Physics and Astronomy, University of California, Davis, 1 Shields Ave, Davis, CA 95616, USA}
\affiliation{Department of Physics, The Chinese University of Hong Kong, Shatin, N.T., Hong Kong SAR, China}

\author[0000-0003-4792-9119]{Ryan L. Sanders}
\affiliation{Department of Physics and Astronomy, University of Kentucky, 505 Rose Street, Lexington, KY 40506, USA}

\author[0000-0001-5984-0395]{Maruša Bradač}
\affiliation{Department of Mathematics and Physics, University of Ljubljana, Jadranska ulica 19, SI-1000 Ljubljana, Slovenia}

\author[0000-0003-4464-4505]{Anthony J. Pahl}
\altaffiliation{Carnegie Fellow}
\affiliation{The Observatories of the Carnegie Institution for Science, 813 Santa Barbara Street, Pasadena, CA 91101, USA}

\author[0000-0003-3509-4855]{Alice E. Shapley}
\affiliation{Department of Physics \& Astronomy, University of California, Los Angeles, 430 Portola Plaza, Los Angeles, CA 90095, USA}

\author[0000-0001-7782-7071]{Richard S. Ellis}
\affiliation{Department of Physics \& Astronomy, University College London, Gower St., London WC1E 6BT, UK}

\author[0000-0001-8426-1141]{Michael W. Topping}\affiliation{Steward Observatory, University of Arizona, 933 N Cherry Avenue, Tucson, AZ 85721, USA}

\author[0000-0001-9687-4973]{Naveen A. Reddy}\affiliation{Department of Physics \& Astronomy, University of California, Riverside, 900 University Avenue, Riverside, CA 92521, USA}

\begin{abstract}
We present a spectral stacking analysis of galaxies at $z\geq6$ observed with the James Webb Space Telescope (JWST). We curate a sample of $64$ galaxies spanning redshifts $z_{\rm spec} = 6.0 - 9.4$ which have NIRSpec medium resolution data. The stacks achieve sufficient signal-to-noise to measure equivalent widths (EW) and velocity centroids ($v_{\rm{cen}}$) of low-ionization species (LIS) absorption features, transmitted Lyman-alpha ({\lya}) emission, and nebular emission lines. Overall, we find our sample has weaker LIS absorption lines ($\rm{EW}(\rm{LIS}) \approx 1~\AA$), smaller $v_{\rm{cen,LIS}} \approx -20 \pm 50~\kms$, and significantly suppressed {\lya} ($\rm{EW}(\rm{Ly\alpha}) \approx 5~\AA$), compared to similar studies undertaken at lower redshift. The weaker LIS absorption may suggest a lower covering fraction of HI and larger escape fraction of ionizing photons from our sample. Additionally, the smaller blueshifted $v_{\rm{cen,LIS}}$ indicates less prevalent or weaker outflows in $z>6$ galaxies. Stacking our sub-sample of \Lya\ emitters (LAEs), we find high EW$(\rm{H}\beta) \approx 170 \pm 4~\AA$ and a detection of nebular \Civ\ emission suggesting higher $\xi_{ion}$ in LAEs at $z>6$. This work showcases the enormous potential for stacked JWST spectra revealing properties of galaxies and their diffuse interstellar medium in the epoch of reionization.
\end{abstract}

\section{Introduction} \label{sec:intro}
The James Webb Space Telescope (JWST) is ushering in a new era for observational studies in the epoch of reionization (EoR) at redshifts $z\gtrsim6$. In less than $3$ years of operation, JWST's Near Infrared Spectograph \citep[NIRSpec;][]{Jakobsen_JWST_NIRSPEC} has had great success confirming candidate galaxies out to $z\sim14$ \citep[e.g.,][]{CurtisLake2023Nature, Bunker_JADES_DR1_spec, ArrabalHaro2023Natur, Carniani2024Natur, Fujimoto2023_CEERS, Harikane2024_SFH_nirspec, Witstok24_z13} while concurrently capturing stunning emission line spectra \citep[e.g.,][]{Bunker_JADES_DR1_spec, Curti2024_mass_Z_SFR_JWST, Cameron2023_nirspec_ism_deep, Sanders23_ionizationpropz2to9, Shapley2023_z3to6}, revealing the physical properties of galaxies in the early Universe. Understanding the characteristics of early galaxies, specifically star-forming galaxies (SFGs) and their interstellar medium (ISM), is crucial for determining their role in reionizing the Universe.

SFGs in the EoR, as opposed to active galactic nuclei (AGN), are thought to be the primary supplier of ionizing photons to the intergalactic medium \citep[IGM; e.g.,][]{Parsa2018_smallAGNsignificanceEoR, Dayal2020, Bouwens2022_Lfunction_z2to9}. Observationally quantifying their contributions to reionization remains a major challenge for extragalactic astronomy. Some on-going efforts include constraining the ionizing production efficiency \citep{Pahl2024_photonproductionefficiency, Matthee23_hb_oiii, Endsley23_nircam_z78, PL23_xion, Simmonds23_xion, Simmonds24_xion, Simmonds24_nircam_xion}, interpreting high-$z$ {\lya} damping wings in galaxy spectra in terms of the IGM neutral fraction \citep{Mason2025_LyaDampWing_JWST}, and comparisons with lower-$z$ analogs leaking Lyman continuum (LyC) flux \citep{Steidel2018_KLCS_z3, SaldanaLopez2022_LzLCS_ISMprops, Pahl2021_LyC_z3, Pahl23_z3_galprops,Flury22_LzLCS_I, Flury22_LzLCS_II}. While these are indirect methods for inferring ionizing contributions, they highlight difficulties with direct observations of SFGs at high-$z$. 

The biggest observational challenge arises from the increasing amounts of intergalactic \Hi\ which attenuates ionizing radiation, making it impossible to directly measure LyC leakage beyond $z\gtrsim 4.5$. Unfortunately, this is well past the end of reionization at $z\gtrsim5.3$ \citep[e.g.][]{Becker2021_MFP, Bosman2022_EoREnds_Z5p3, Spina2024_LyaDampWing, Zhu_2024_dampwing_hydroisland}, preventing any observation of LyC photons within the EoR. Furthermore, efforts to probe the ISM and massive star properties \citep[e.g.,][]{Shapley2003_z3LBG, Steidel2010_cgmkinematics, Steidel2016_nebspec, Leitherer2011,Senchyna2019} are generally possible only for uncharacteristically bright galaxies in the EoR, or for rare sources which are strongly gravitationally lensed (e.g., \citealt{Pettini2002, Jones2013_z4Lensed, Leethochawalit2016_fesc} at $z\approx3$--4, and \citealt{Topping2024_CIV_highz,Boyett2024_z9_lensed} at $z>6$). 

Spectral stacking can offer an alternative way to study the rest-UV properties of the fainter, but more abundant, galaxy population. This method takes a statistical average across a sample of sources with moderate signal-to-noise ratio (SNR). In this way, the individual probes and associated scatter are traded for detectability of averaged spectral features in a composite spectrum. 
Recent work has shown the promise of JWST stacking analyses of high-$z$ galaxies using NIRSpec medium resolution and PRISM data \citep[e.g.,][]{Hu2024_ISM_stacks, Borsani2024ApJ_R100_stacks}. In particular, \citet{Borsani2024ApJ_R100_stacks} used rest-frame optical features to infer galaxy properties (e.g., ages, masses, metallically of stellar populations, and ISM conditions) at $z\geq5$. This work follows several studies at lower redshifts $z\simeq2$--5 \citep[e.g.,][]{Shapley2003_z3LBG, Jones2013_z4Lensed, Pahl2020_z5stacks, Du2018ApJ_specprops_LBGs}, which have characterized ISM absorption properties using rest-UV absorption lines in composite spectra. The low-ionization species (LIS) absorption lines are especially notable as they trace the neutral ISM (i.e., \Hi) and strongly correlate with \Hi\ \Lya\ emission. 

Metal LIS lines hold great promise for study in the EoR given their detectability with JWST at $z>6$ and their physical association with the surrounding neutral hydrogen \citep[e.g.,][]{Trainor2015_escapeinggas, RiveraThorsen2017_ISM, Bhagwat2022_cospatial, Gazagnes2018A&A_neutral, Reddy2016ApJ_gascovfrac, Chisholm2018_fesc_predictions}. Equivalent widths (EW) from saturated LIS lines can provide insight into the low-ionization covering fraction {\fcovlis}. The {\fcovlis} has been shown to positively correlate with the \Hi\ covering fraction {\fcovhi}, which in turn regulates the ionizing escape fraction {\fesc} \citep[e.g.,][]{Chisholm2018_fesc_predictions}. While the correlation between {\fcovlis} and {\fcovhi} is not one-to-one \citep{Reddy2016ApJ_gascovfrac}, the accessibility of LIS lines and their physical association with \Hi\ and {\fesc} makes them a valuable means to probe high-$z$ galaxy contributions to reionization. Further evidence for a physical link between LIS absorption and \Hi\ can be found in the anti-correlation between EW(LIS) and Lyman~$\alpha$ ({\lya}) emission strength \citep[e.g.,][]{Shapley2003_z3LBG, Du2018ApJ_specprops_LBGs, Jones2013_z4Lensed, Reddy2016ApJ_gascovfrac,Chisholm2018_fesc_predictions, Pahl2020_z5stacks}. Given the success of JWST/NIRSpec stacking analyses at $z \geq5$ and the well-studied LIS line diagnostics at low-$z$, our objective in this work is to stack JWST medium resolution spectra to investigate the interstellar absorption features from galaxies that significantly impact the state of the IGM. 

In this study, we present a spectral stacking analysis performed on $64$ $z\geq6$ galaxies observed with JWST/NIRSpec medium resolution spectroscopy ($R\sim 1000$). In Section~\ref{sec:PANCAKEZ_sample} we discuss the spectra and the selection criteria applied to refine the PANCAKEZ sample of galaxies. In Section \ref{sec:stacking_PANCAKEZ} we describe how composite stacks were constructed using the finalized PANCAKEZ sample. In Section \ref{sec:Measurements} we explain how spectral features were isolated and measured from the composite spectra. In Section \ref{sec:discussion} we compare our stacks to lower-$z$ studies and discuss the implications of our results. We summarize our conclusions in Section \ref{sec:conclusion}. Throughout this paper we assume a $\Lambda$CDM concordance cosmology with $\Omega_m = 0.3 $, $\Omega_{\Lambda}= 0.7 $ and Hubble constant $H_{0} =  70~\rm{km\,s}^{-1} \, \rm{Mpc}^{-1}$.

\section{The PANCAKEZ Sample} \label{sec:PANCAKEZ_sample}
The main goal of this study is to characterize the interstellar medium of $z>6$ SFGs by creating composite spectra, with a particular focus on measuring rest-UV absorption lines. To that end, we require galaxies that have rest-UV spectral coverage with the NIRSpec medium resolution ($R\sim 1000$) gratings and are spectroscopically confirmed to be at $z_{\rm{spec}}\geq 6$. Additionally, individual spectra must achieve sufficient SNR in their rest-frame UV continuum to ensure good sensitivity to absorption lines. With these requirements in mind, we compiled and curated an optimum set of galaxies that make up the PANCAKEZ sample. We provide a summary of the selection process in Table \ref{tab:PancakezSampleTable}. 

\subsection{Sample selection}

\subsubsection{Surveys}\label{sec:surveys}
The galaxy spectra used in this work originate from three JWST surveys: the JWST Advanced Deep Extragalactic Survey \citep[JADES DR3;][]{JADES_2023Rieke, Bunker_JADES_DR1_spec}, the Cosmic Evolution Early Release Science Survey \citep[CEERS;][]{Arrabal2023_CEERS_spec, Finkelstein2023_CEERS}, and Assembly of Ultradeep Rest-optical Observations Revealing Astrophysics \citep[AURORA;][]{Shapley2024_AURORA_Survey, Sanders2024_AURORA}. Both JADES\footnote{\url{https://archive.stsci.edu/hlsp/jades}} and CEERS\footnote{\url{https://ceers.github.io/releases.html}} reduced data are publicly available. The number of galaxies observed spectroscopically with all three surveys totals over $4000$. We formed our initial sample by requiring galaxies to have rest-UV $R\sim1000$ medium resolution spectra (i.e., observations with the G140M grating; our targets additionally have spectra in the G235M and G395M gratings although this is not a strict requirement for our analysis). We removed any objects for which there is published evidence for the presence of an AGN. We require a spectroscopic confirmation of $z_{\rm{spec}} \geq 6$, either through JWST observations or previous studies. These cuts formed the initial parent sample of $223$ galaxies (Table~\ref{tab:PancakezSampleTable}). 

\subsubsection{Spectral coverage} \label{sec:JWST_spec_coverage}

Within the redshift range ($6<z<9.4$) of this sample, the NIRSpec gratings G140M and G395M are most important for our analysis. The G140M grating has a rest-frame spectral coverage of $\sim 850-2700~\mathrm{\AA}$ depending on the target redshift, which covers the LIS absorption features of interest. The G395M grating covers $\sim 2900 - 7500~\mathrm{\AA}$, which includes the rest-optical nebular emission lines used to determine systemic redshifts. The vast majority of the parent sample, with the exception of 12 galaxies, has G235M coverage as well. After additional filtering (see below), the final stack contained only two galaxies (JADES-$00099302$ and JADES-$00265801$) without G235M coverage, which has negligible impact on our analysis. 

\subsubsection{Spectroscopic redshifts} \label{sec:spectroscopic_redshifts}
Constructing composite spectra requires accurate systemic redshifts. We remeasured spectroscopic redshifts of all $223$ galaxies using H$\beta$, \Oiii$4960$, and \Oiii$5008$ nebular emission lines. We visually inspected each galaxy's spectrum with an interactive spectral analysis tool, \textsc{XTRIMPY}\footnote{Public github repository: \url{https://github.com/yuguangchen1/xtrimpy}} and fit a Gaussian profile to each detected emission line. Using the observed central wavelength from the fits, an average redshift was calculated when at least $2$ out of the $3$ emission lines were present. From the parent sample of $223$, we re-measured $z_{\rm{spec}}$ with nebular lines for $131$ galaxies. We were unable to confirm the  $z_{\rm{spec}}$ from $92$ galaxies, which exhibited no obvious nebular emission or other significantly detected features in their medium-resolution spectra.This ultimately resulted in their removal from the working sample. While $27$ of these $92$ galaxies had high-confidence spectroscopic redshifts, a composite of their $R\sim1000$ spectra did not show strong detection of spectral features, and only a minority ($10$ out of these $27$) satisfy our SNR threshold discussed in Section~~\ref{continuumSNRthreshold_sec}. We thus proceeded with those $131$ galaxies for which we confirmed robust redshifts from medium-resolution spectra.

\begin{deluxetable}{lcccc}
\tabletypesize{\scriptsize}
\tablewidth{0pt} 
\tablecaption{The PANCAKEZ Sample \label{tab:PancakezSampleTable}}
\tablehead{\colhead{Survery} & \colhead{\makecell{Galaxies with \\ $z_{\rm{spec}}>6$}} & \colhead{\makecell{Galaxies with \\ H$\beta +[\mathrm{O\,III}]$}} & \colhead{\makecell{Galaxies with \\ $\rm{SNR_{\rm{cont}}}\geq\rm{SNR}_{thresh}$}} \\
}
\startdata 
JADES & 175 & 98 & 46 \\ 
CEERS & 35 & 22 & 9 \\
AURORA & 13 & 11 & 9 \\
\hline
Total & 223 & 131 & 64 \\ 
\enddata
\tablecomments{The number of galaxy spectra after applied selection criteria.}
\end{deluxetable}

\subsubsection{Continuum signal-to-noise ratio} \label{continuumSNRthreshold_sec}
From the $131$ galaxies with robust spectroscopic redshifts, we measured the individual median continuum SNRs directly using the rest-frame wavelength range from $1450 - 1500$~\AA. This was chosen as a relatively featureless region for SFGs. A subset of galaxies ($10$) had no spectral coverage in this wavelength range due to the chip gap between the two NIRSpec detectors; we refer to these as chip-gap galaxies (CGGs). To include these cases, we adopted a secondary median continuum SNR for all galaxies in a spectral range ($1265 - 1300$~\AA) which had guaranteed spectral coverage in the CGGs. 

\begin{figure}
\centering
\includegraphics[width=\columnwidth]{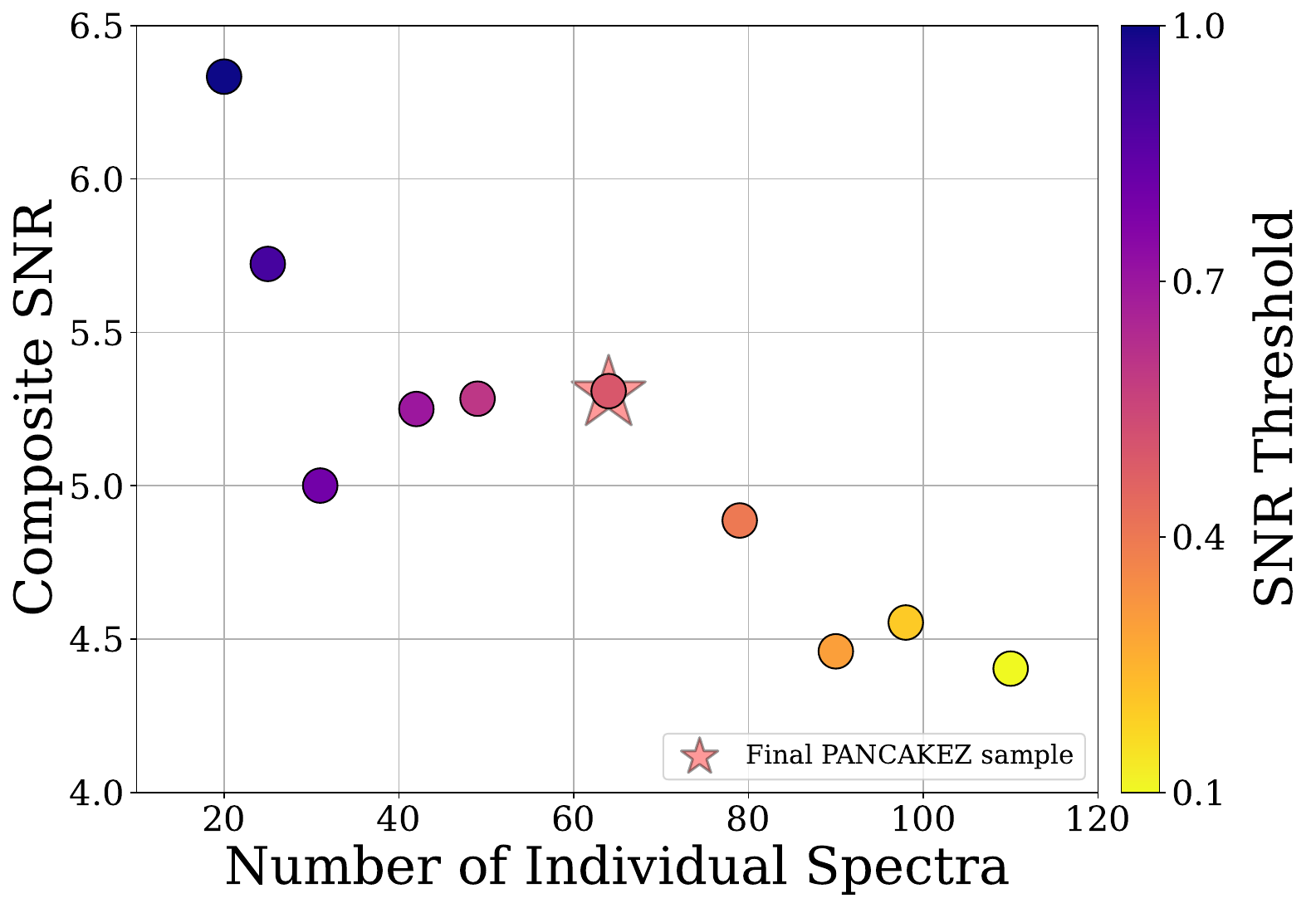}
\caption{The composite spectrum continuum SNR as a function of the number of galaxies combined to create the stack. The color of each symbol represents the minimum continuum SNR (SNR Threshold, or $\rm{SNR}_{thresh}$) that individual galaxies must achieve in order to be included. As expected for our methodology, the highest composite SNR ($\approx 6.3$) is achieved by combining galaxies with relatively high individual SNR, but at the cost of having a small sample size ($N = 20$ out of the parent sample of $131$). Conversely, combining galaxies with low individual continuum SNR results in having a lower SNR in the stacked composite spectrum (yellow circle). The star marker indicates the fiducial $N=64$ PANCAKEZ sample, which balances the desires of a large representative sample and good continuum SNR.}
\label{fig:SNR_histogram}
\end{figure}

For our stacking analysis we seek to capture the average properties of a large sample, with sufficient SNR to detect strong absorption features. Adding large numbers of faint galaxies to the sample is detrimental to the SNR, while restricting to the brightest sources could lead to results which are not representative of the broader population. 
We investigated the tradeoff between sample size and SNR by constructing a series of composite spectra (see Section~\ref{sec:stacking_PANCAKEZ} for details of the stacking method) with different input SNR thresholds ($\rm{SNR}_{thresh}$). For each composite, we include the subset of galaxies with continuum $\rm{SNR} \geq \rm{SNR}_{thresh}$ (measured from rest-frame $1450 - 1500$~\AA). We then measure the SNR of the composite spectrum in the same wavelength range. 
Figure~\ref{fig:SNR_histogram} shows how the composite SNR and stacking sample size ($N$) fluctuate as the quality of the input spectra (i.e., SNR threshold) is varied. 
Increasing the sample size beyond the $N=20$ highest-SNR sources results in progressively lower SNR in the composite spectra. However, the effect is modest, such that tripling the input sample results in only a $\sim$15\% reduction in SNR. 
We thus chose a SNR threshold of $0.5$ (red star in Figure~\ref{fig:SNR_histogram}) to maintain a sufficiently high rest-UV continuum SNR ($>5$ in the composite spectrum) for robust absorption line analysis, while maintaining a reasonably large sample size to avoid being dominated by the brightest galaxies in our sample.
This results in a final PANCAKEZ stacking sample of $64$ galaxies. 

\subsection{Sample statistics}

\begin{figure}
\centering
\includegraphics[width=\columnwidth]{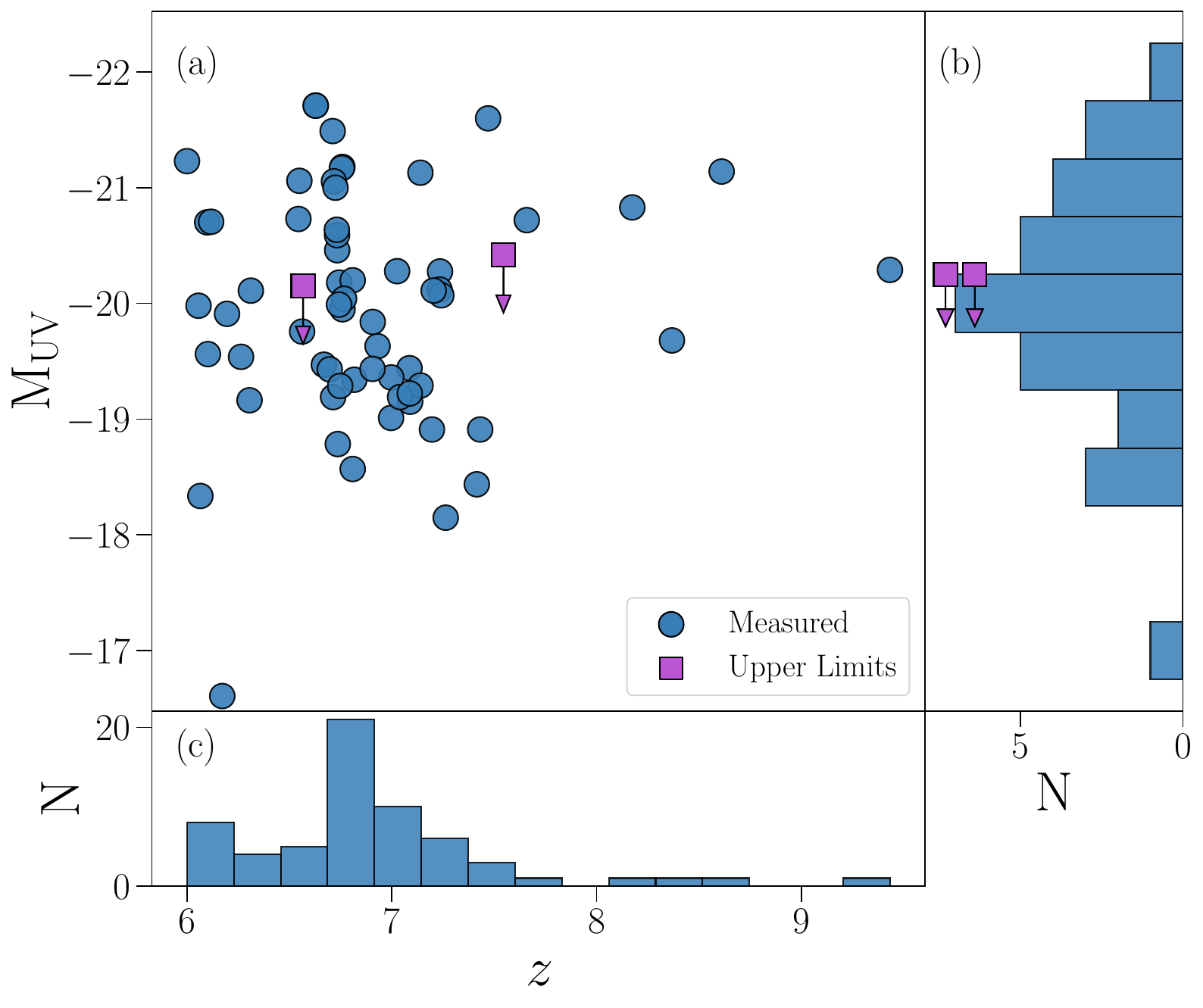}
\caption{The overall distribution of $\rm{M}_{\rm{UV}}$ and redshift for the $64$ galaxies that make up the final PANCAKEZ sample. \textbf{(a):} The distribution of $\rm{M}_{\rm{UV}}$ as a function of redshift. \textbf{(b):} The number distribution of  $\rm{M}_{\rm{UV}}$. \textbf{(c):} The number of galaxies in a given redshift bin. $\rm{M}_{\rm{UV}}$ upper limits for two galaxies are shown as purple squares with arrows. For reference, the characteristic UV luminosity is $M^{*}_{\rm{UV}} = -21.15 \pm0.13$ at $z \approx 6.8$ as reported by \cite{Bouwens2021_LF_z29}.}
\label{fig:M_UV_histogram_64gals}
\end{figure}

Properties of the 64 galaxies which comprise the PANCAKEZ stacking sample are given in Table~\ref{tab:PANCAKEZ_properties}. The redshift distribution of the sample has a mean and sample standard deviation of $\left<z_{\rm{spec,64}}\right> = 6.89 \pm 0.61$. 

\begin{deluxetable*}{rcrrcccc}
\tabletypesize{\scriptsize}
\tablecaption{PANCAKEZ Sample Properties \label{tab:PANCAKEZ_properties}} 
\tablewidth{0.3\textwidth}
\tablehead{ \colhead{Galaxy ID} & \colhead{Survey} & \colhead{R.A(deg)} & \colhead{Dec(deg)} & \colhead{$z_{\rm{spec}}$} & \colhead{LAE} & \colhead{$\rm{M}_{\rm{UV}}$} & \colhead{Reference}
}
\startdata
{$100182$} & AURORA & {$189.17717$} & {$62.29152$} & {$6.7317$} & No & {$-20.64$} & [$1$] \\
{$425968$} & AURORA & {$150.14514$} & {$02.22618$} & {$6.7364$} & No & {$-18.78$} & [$1$] \\
{$100164$} & AURORA & {$189.13914$} & {$62.27530$} & {$6.7423$} & No & {$-19.99$} & [$1$] \\
{$100163$} & AURORA & $189.13807$ & $62.27444$ & $6.7479$ & No & $-19.27$ & [$1$] \\
{$419213$} & AURORA & $150.14140$ & $02.21142$ & $6.8090$ & No & $-18.57$ & [$1$] \\
{$100167$} & AURORA & $189.20260$ & $62.27552$ & $6.9060$ & No & $-19.43$ & [$1$] \\
{$100159$} & AURORA & $189.13759$ & $62.26520$ & $7.0405$ & No & $-19.19$ & [$1$] \\
{$100174$} & AURORA & $189.19571$ & $62.28242$ & $7.0878$ & No & $-19.22$ & [$1$] \\
{$100026$} & AURORA & $189.22512$ & $62.28629$ & $7.2043$ & Yes & $-20.11$ & [$1$] \\
\hline
{$1345\_397$} & CEERS & $214.83620$ & $ 52.88269$ & $6.0001$ & No & $-21.23$ & [$2$] \\
{$1345\_603$} & CEERS & $214.86725$ & $ 52.83674$ & $6.0563$ & No & $-19.98$ & [$2$] \\
{$1345\_1561$} & CEERS & $215.16610$ & $53.07076$ & $6.1954$ & No & $-19.91$ & [$2$] \\
{$1345\_1115$} & CEERS & $215.16282$ & $53.07310$ & $6.2986$ & No & $-$ & [$2$] \\
{$1345\_1160$} & CEERS & $214.80505$ & $52.84588$ & $6.5669$ & No & $>-20.15$ & [$2$] \\
{$1345\_698$} & CEERS & $215.05032$ & $ 53.00744$ & $7.4702$ & Yes & $-21.6$ & [$2$] \\
{$1345\_689$} & CEERS & $214.99905$ & $ 52.94198$ & $7.5453$ & No & $>-20.42$ & [$2$] \\
{$1345\_1149$} & CEERS & $215.08971$ & $52.96618$ & $8.1742$ & No & $-20.83$ & [$2$] \\
{$1345\_1029$} & CEERS & $215.21876$ & $53.06986$ & $8.6107$ & No & $-21.14$ & [$2$] \\
\hline
{$99302$} & JADES & $53.12588$ & $-27.81823$ & $6.0652$ & No & $-18.34$ & [$1$], [$5$] \\
{$9877$} & JADES & $53.15952$ & $-27.77152$ & $6.0986$ & No & $-20.70$ & [$1$], [$5$] \\
{$209277$} & JADES & $53.15618$ & $-27.77576$ & $6.1022$ & No & $-19.56$ & [$1$], [$5$] \\
{$30141885$} & JADES & $53.15952$ & $-27.77152$ & $6.1178$ & No & $-20.71$ & [$1$], [$5$] \\
{$9547$} & JADES & $53.20799$ & $-27.78994$ & $6.1723$ & No & $-16.61$ & [$1$], [$5$] \\
{$58850$} & JADES & $53.09517$ & $-27.76062$ & $6.2639$ & No & $-19.96$ & [$3$] \\
{$9597$} & JADES & $53.16611$ & $-27.77204$ & $6.3055$ & Yes & $-19.16$ & [$1$], [$5$] \\
{$988$} & JADES & $189.16214$ & $62.26381$ & $6.3117$ & Yes & $-20.11$ & [$1$], [$5$] \\
{$57330$} & JADES & $189.22885$ & $62.20400$ & $6.5441$ & No & $-20.73$ & [$4$] \\
{$78891$} & JADES & $189.22582$ & $62.20421$ & $6.5488$ & No & $-21.06$ & [$4$] \\
{$1967$} & JADES & $189.16501$ & $62.30018$ & $6.5620$ & No & $-19.76$ & [$1$], [$5$] \\
{$58930$} & JADES & $53.10538$ & $-27.72347$ & $6.6285$ & No & $-21.71$ & [$4$] \\
{$18533$} & JADES & $189.12122$ & $62.28641$ & $6.6683$ & No & $-19.47$ & [$4$] \\
{$39799$} & JADES & $189.26354$ & $62.15479$ & $6.6971$ & No & $-19.43$ & [$4$] \\
{$38428$} & JADES & $189.17928$ & $62.27590$ & $6.7113$ & No & $-21.49$ & [$4$],  \\
{$2113$} & JADES & $189.17033$ & $62.22950$ & $6.7134$ & No & $-19.19$ & [$4$] \\
{$38432$} & JADES & $189.18617$ & $62.27086$ & $6.7175$ & No & $-21.06$ & [$1$], [$5$] \\
{$89464$} & JADES & $189.18658$ & $62.27090$ & $6.7250$ & No & $-21.00$ & [$4$] \\
{$38420$} & JADES & $189.17512$ & $62.28226$ & $6.7327$ & Yes & $-20.59$ & [$4$] \\
{$1972$} & JADES & $189.18840$ & $62.30305$ & $6.7333$ & No & $-20.46$ & [$1$], [$5$] \\
{$1076$} & JADES & $189.1385$ & $62.27561$ & $6.7430$ & No & $-20.18$ & [$4$] \\
{$926$} & JADES & $189.07981$ & $62.25644$ & $6.7562$ & No & $-21.17$ & [$4$] \\
{$896$} & JADES & $189.08266$ & $62.25247$ & $6.7589$ & No & $-21.18$ & [$4$] \\
{$954$} & JADES & $189.15197$ & $62.25963$ & $6.7596$ & Yes & $-19.95$ & [$1$], [$5$] \\
{$73977$} & JADES & $189.18550$ & $62.17981$ & $6.7668$ & No & $-20.04$ & [$4$] \\
{$18536$} & JADES & $189.15531$ & $62.28647$ & $6.8091$ & Yes & $-20.20$ & [$4$] \\
{$9104$} & JADES & $189.24527$ & $62.25253$ & $6.8168$ & No & $-19.34$ & [$4$] \\
{$1075$} & JADES & $189.20260$ & $62.27551$ & $6.9070$ & Yes & $-19.84$ & [$4$] \\
{$13609$} & JADES & $53.11730$ & $-27.76409$ & $6.9305$ & No & $-19.63$ & [$4$] \\
{$7424$} & JADES & $189.23290$ & $62.24738$ & $6.9965$ & No & $-19.01$ & [$4$] \\
{$2316$} & JADES & $189.16254$ & $62.25824$ & $6.9971$ & No & $-19.36$ & [$4$] \\
{$1931$} & JADES & $189.06964$ & $62.28102$ & $7.0261$ & No & $-20.28$ & [$1$], [$5$] \\
{$1129$} & JADES & $189.17980$ & $62.28240$ & $7.0865$ & Yes & $-19.44$ & [$4$] \\
{$13041$} & JADES & $189.20377$ & $62.26843$ & $7.0898$ & Yes & $-19.15$ & [$4$] \\
{$24819$} & JADES & $189.13649$ & $62.22340$ & $7.1400$ & No & $-21.13$ & [$4$] \\
{$66336$} & JADES & $189.25929$ & $62.23546$ & $7.1404$ & Yes & $-19.29$ & [$4$] \\
{$13905$} & JADES & $53.11833$ & $-27.76901$ & $7.1960$ & No & $-18.91$ & [$4$] \\
{$11547$} & JADES & $53.16483$ & $-27.78826$ & $7.2332$ & No & $-20.12$ & [$4$] \\
{$30099449$} & JADES & $53.16170$ & $-27.78540$ & $7.2355$ & No & $-20.28$ & [$1$], [$5$] \\
{$15423$} & JADES & $53.16958$ & $-27.73806$ & $7.2420$ & No & $-20.07$ & [$4$] \\
{$20046866$} & JADES & $53.18405$ & $-27.79783$ & $7.2633$ & No & $-18.15$ & [$1$], [$5$] \\
{$4685$} & JADES & $189.09630$ & $62.23914$ & $7.4152$ & No & $-18.44$ & [$1$], [$5$] \\
{$60331$} & JADES & $189.27524$ & $62.21244$ & $7.4317$ & No & $-18.91$ & [$4$] \\
{$12637$} & JADES & $53.13347$ & $-27.76039$ & $7.6591$ & Yes & $-20.59$ & [$3$] \\
{$45131$} & JADES & $189.21139$ & $62.17030$ & $8.3681$ & No & $-19.68$ & [$4$] \\
{$265801$} & JADES & $53.11244$ & $-27.774620$ & $9.4326$ & No & $-20.29$ & [$4$] \\
\enddata
\tablecomments{References: [$1$] this work, [$2$] \citet{Nakajima2023ApJ_CEERS}, [$3$] 
\citet{Saxena2024AA_JADES}, [$4$] \citet{Tang2024_Lya_SFGs_z613}, [5] \citet{Rieke2023}}
\end{deluxetable*}

To assess the overall population of galaxies that make up the final PANCAKEZ sample, we plot the distribution of absolute rest-UV magnitudes $\rm{M}_{\rm{UV}}$ in Figure \ref{fig:M_UV_histogram_64gals}. In general these are based on observed JWST/NIRCam F150W photometry, corresponding to $\sim$2000~\AA\ in the rest frame of our sample. 
For objects taken from the CEERS survey, we use published $\rm{M}_{\rm{UV}}$ values from \cite{Nakajima2023ApJ_CEERS} with the exception of one galaxy which is not reported (CEERS-$1345 \textunderscore 1115$; we exclude this galaxy from Figure~\ref{fig:M_UV_histogram_64gals} and the mean $\rm{M}_{\rm{UV}}$ values). $30$ target galaxies from the JADES survey also had previously recorded $\rm{M}_{\rm{UV}}$ values \citep{Tang2024_Lya_SFGs_z613, Saxena2024AA_JADES} which we adopt here. For the remaining $16$ galaxies from JADES we calculate $\rm{M}_{\rm{UV}}$ using the available Kron \citep{kron80} photometry of F150W imaging from \citet{Rieke2023} (except JADES-$58850$ which lacks F150W, where we instead use F182M). The $\rm{M}_{\rm UV}$ values for AURORA galaxies are calculated using photometric catalogs taken from the DAWN JWST Archive,\footnote{\url{https://dawn-cph.github.io/dja/}} \citep{Valentino2023_DJA} which contains HST and NIRCam photometry. For $4$ AURORA galaxies lacking F150W photometry, we used HST/WFC3 F160W instead. 
We report an average $\rm{M}_{\rm{UV}} = -19.7$ with a sample standard deviation of $1.1$ from the $61$ PANCAKEZ objects which have a recorded $\rm{M}_{\rm{UV}}$ (excluding the two upper-limit values from CEERS-$1345 \textunderscore 1160$, and CEERS-$1345 \textunderscore 689$).

\section{Stacking PANCAKEZ}\label{sec:stacking_PANCAKEZ}

The final set of $64$ galaxies are combined to produce the PANCAKEZ composite spectra analyzed in this work. This section describes the stacking methodology along with the specifications that differentiate stacks from one another. 

\subsection{Building the composite} \label{sec:PANCAKEZ_fullComposite}

\begin{figure*}
\centering
\includegraphics[width=\textwidth]{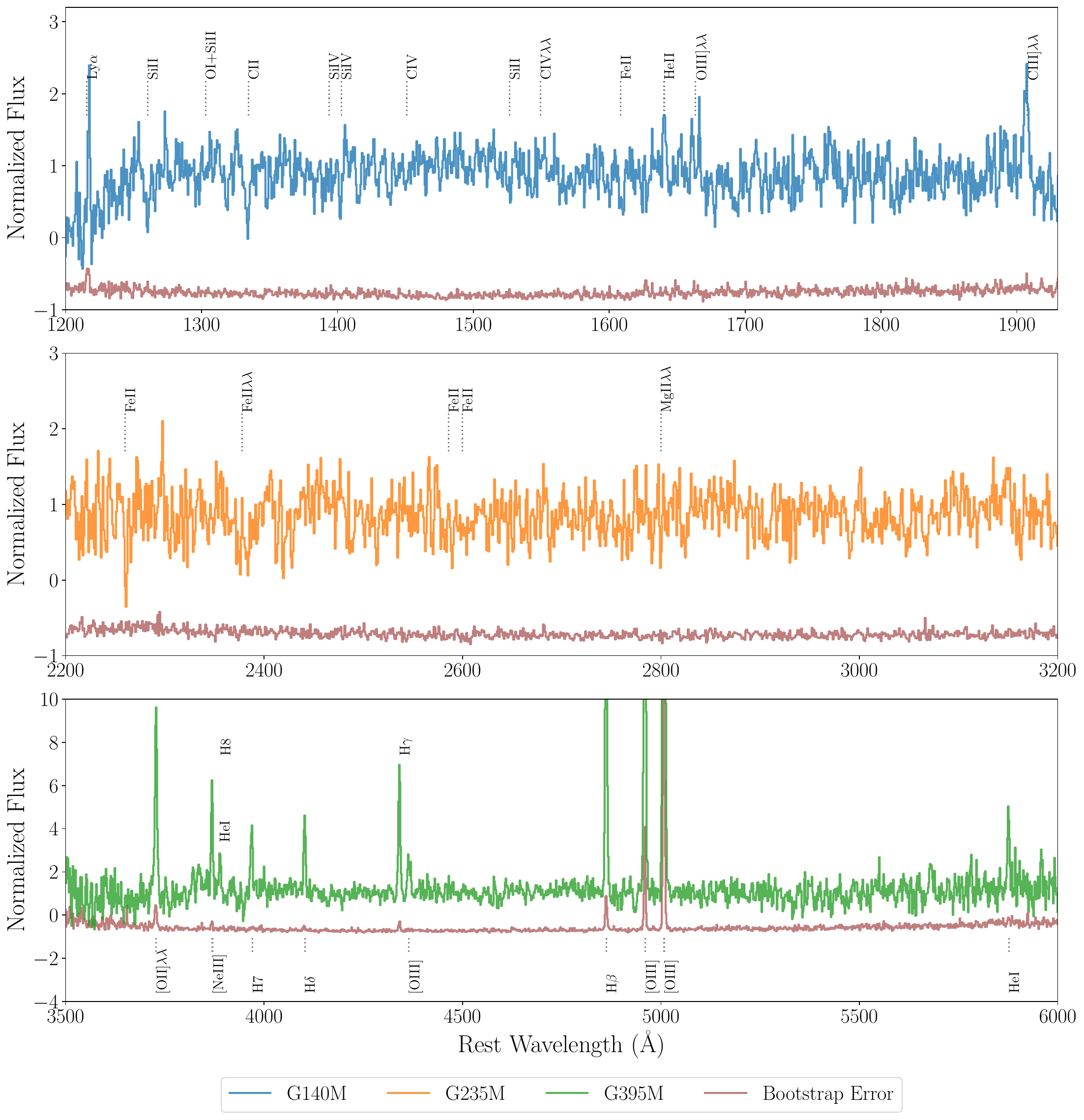}
\caption{The full PANCAKEZ composite spectrum for each JWST grating: G140M (blue), G235M (orange), and G395M (green). The filter composites are constructed by averaging normalized individual spectra in units of $f_{\nu}$ (Section~\ref{sec:stacking_PANCAKEZ}). Prominent spectral features are labeled with names and dotted vertical lines. Closely separated doublets are marked with $\lambda \lambda$. The bootstrap error spectrum for each filter stack is shown in maroon under the composite, subtracted uniformly by 1 for clarity. This figure showcases the quality of our composite spectrum constructed from our fiducial sample of $64$ $z>6$ galaxies.} 
\label{fig:Full-IndividualFilterComposites}
\end{figure*}

\begin{figure*}
\centering
\includegraphics[width=\textwidth]
{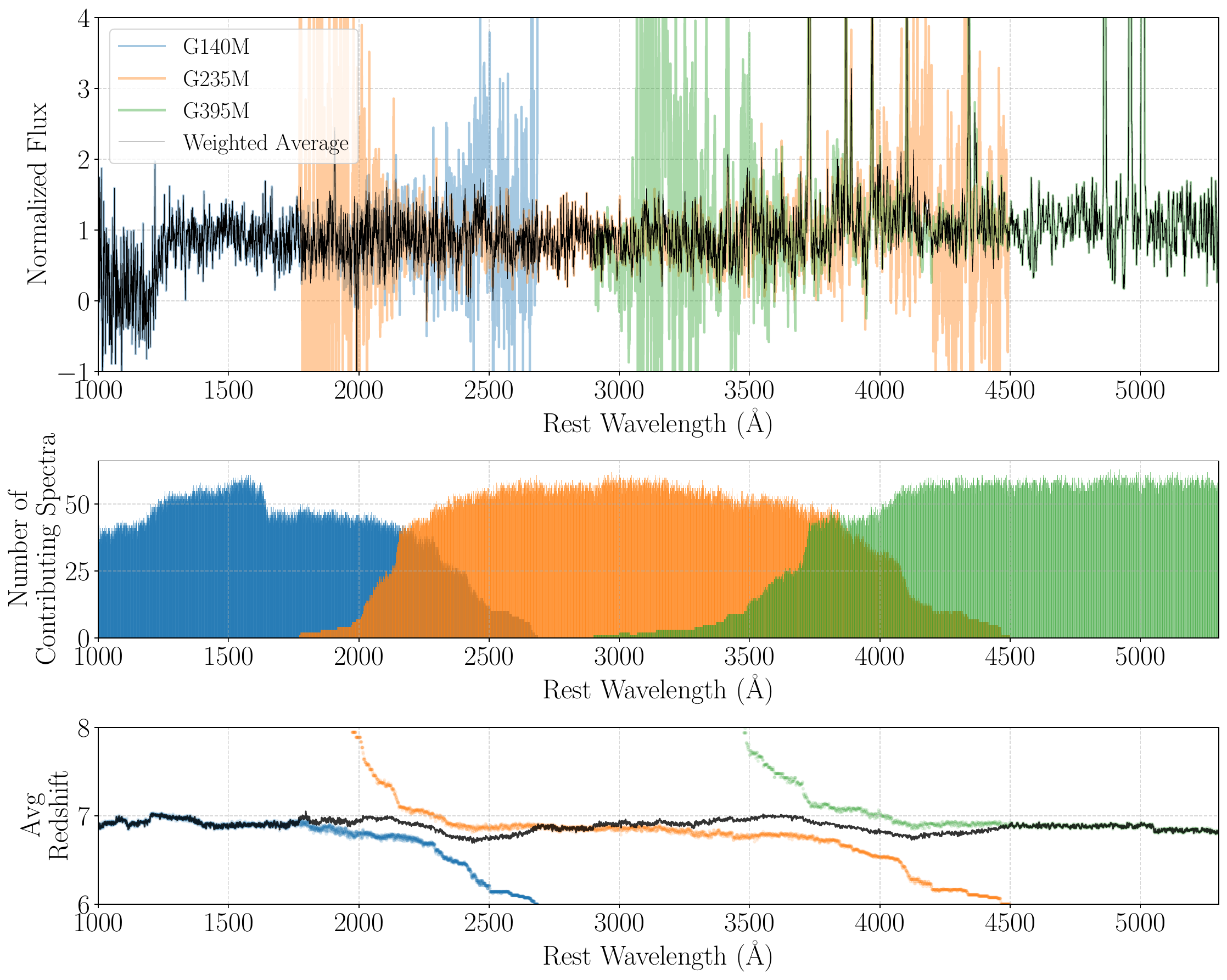}
\caption{\textbf{Top:} The full composite spectrum of the $64$ $z_{\rm{spec}}\geq6$ galaxies. \textbf{Middle:} The number of spectra combined at each wavelength, shown for each filter individually. \textbf{Bottom:} The average redshift calculated at each wavelength of the combined galaxies. Black represents the weighted average. Overall our fiducial stack probes a consistent sample of galaxies across a broad wavelength range in each filter composite, and across the full wavelength range of the weighted average. The majority of rest-frame UV ISM features used in this work are captured entirely by the G140M filter, while rest-frame optical emission lines are predominantly covered by the G395M filter.}
\label{fig:Full-Composite_SpectrumAllFilter_NumberHistogram_AvgRedshifts}
\end{figure*}

Using the nebular systemic redshifts (Section~\ref{sec:spectroscopic_redshifts}), we shifted all spectra to their respective rest frames. We interpolated the spectra, in $f_{\nu}$ flux density units, to a common wavelength grid such that the individual spectra are Nyquist sampled. Before stacking, individual spectra are normalized to their continuum levels measured from $1450 - 1500~\mathrm{\AA}$. This range is chosen because it is mostly free of strong spectral line features while remaining close to the rest-UV absorption lines of greatest interest for this work. For the handful of CGGs without the $1450 - 1500~\mathrm{\AA}$ coverage (see Sec \ref{continuumSNRthreshold_sec}), the spectra were normalized by fitting a simple power law to the continuum spectra around the chip gap and inferring the $1450-1500~\mathrm{\AA}$ continuum level from the best fit. 

The composite spectra are constructed using a $3\sigma$-clipped mean of normalized individual spectra. The sigma-clipped mean is commonly used and has been shown to be representative of the sample's overall characteristics. The results are nearly identical to composite spectra constructed using a mean with min/max rejection \citep[e.g.,][]{Shapley2003_z3LBG,Jones2012_faintLBG}, such that they can be reliably compared with the PANCACKEZ composite. This method is also close to the median stack, while optimizing the SNR by rejecting outliers \citep{cheny20}. Uncertainties in the composite spectra are estimated from bootstrap resampling. This process generates $2000$ composite spectra. Each spectrum is generated from the same sigma-clipped method mentioned above with individual galaxies randomly drawn from the fiducial sample.

\begin{figure*}
\centering
\includegraphics[width=\textwidth]{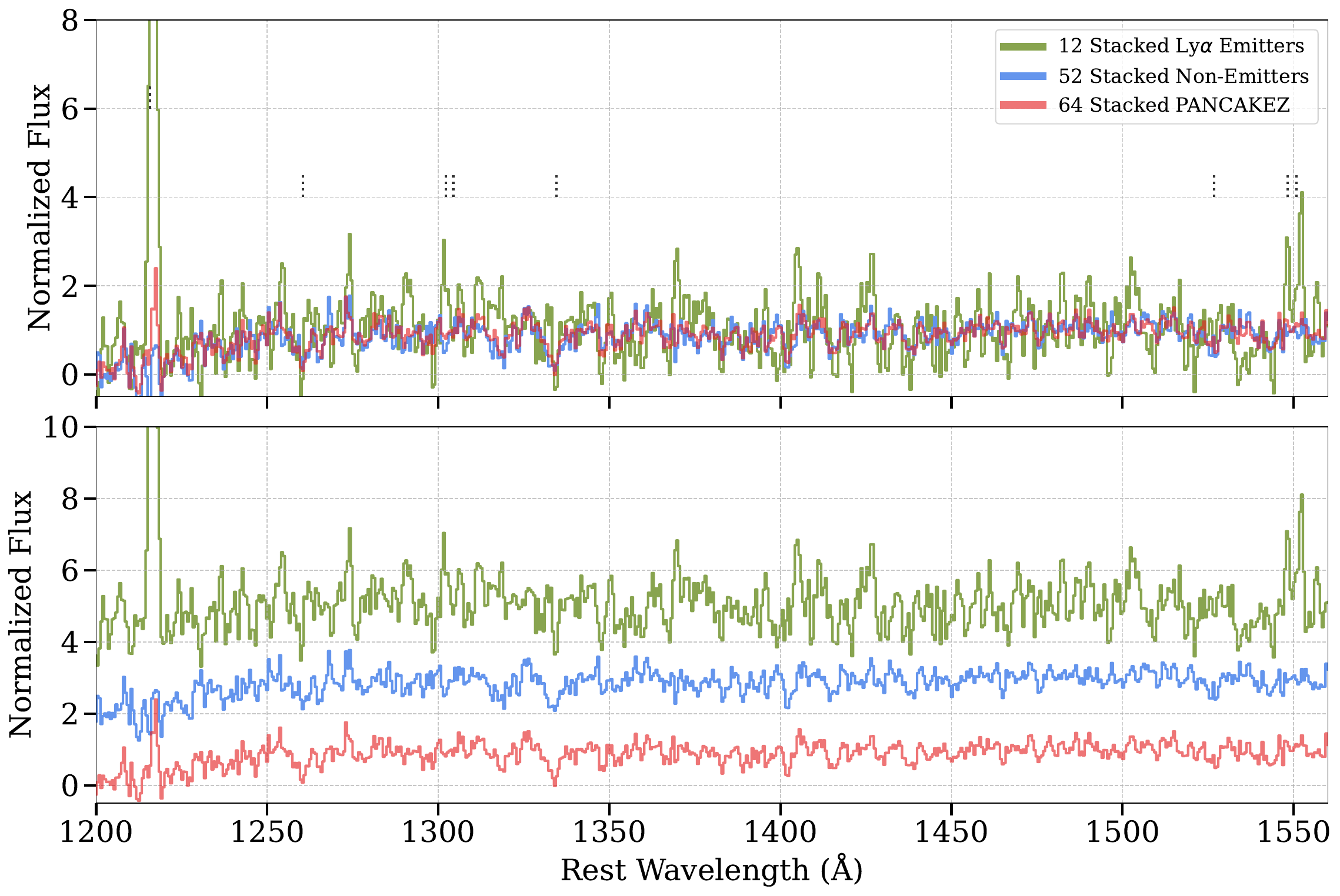}
\caption{The rest-UV composite spectra produced from the full $64$-galaxy PANCAKEZ sample (red), our bin of $12$ LAEs (green), and our bin of $52$ non-LAEs (blue). The top panel displays all three composites plotted directly top of each other while the bottom panel has them vertically separated by increments of $2$. Black dotted lines denote various spectral features of interest (see Figure~\ref{fig:Full-IndividualFilterComposites}): \Hi$\lambda1216$ (\Lya), \Siiiabs$\lambda 1260$, \Oiabs$\lambda 1302$, \Siiiabs$\lambda 1304$, \Ciiabs$\lambda 1334$, \Siiiabs$\lambda 1526$, and \Civ $\lambda \lambda 1548,1550$.}
\label{fig:restUV_LAE_nonLAE_comparisons}
\end{figure*}

Since most analyses in this work are performed independently at rest-UV (covered by G140M) and rest-optical (G395M) wavelengths, composite spectra are constructed independently for each grating. In Figure~\ref{fig:Full-IndividualFilterComposites} we display the individual grating composites for our full $64$-galaxy sample. Relevant spectral features are labeled and the bootstrap error is shown in maroon. 

The full wavelength coverage of the $3$ gratings is presented in Figure \ref{fig:Full-Composite_SpectrumAllFilter_NumberHistogram_AvgRedshifts}, which additionally shows the number of spectra contributing to each wavelength bin (middle panel of Figure~\ref{fig:Full-Composite_SpectrumAllFilter_NumberHistogram_AvgRedshifts}) and the average redshift of the contributing galaxies (bottom panel of Figure~\ref{fig:Full-Composite_SpectrumAllFilter_NumberHistogram_AvgRedshifts}). Consistency in both the number of contributing spectra and the average redshift implies the grating composites probe the same set of galaxies, across nearly the full wavelength range. Substantial deviation occurs only at the edges of the grating wavelength ranges, with shorter wavelengths covered by higher-$z$ targets (and vice versa). 

We also perform a weighted average of the grating spectra in the regions of rest-frame wavelength overlap (black lines in Figure~\ref{fig:Full-Composite_SpectrumAllFilter_NumberHistogram_AvgRedshifts}). Weights are assigned by the number of spectra contributing towards the stack (see middle panel of Figure~\ref{fig:Full-Composite_SpectrumAllFilter_NumberHistogram_AvgRedshifts}) at each given wavelength bin. This resulting full composite spectrum probes the full sample nearly uniformly across a large spectral range from 1000 to $>5500$~\AA. This allows for robust simultaneous measurements of LIS absorption features and nebular emission lines. In both the weighted average and individual grating composite spectra, we see prominent LIS absorption features such as \Siiiabs{}~$\lambda1260$ and \Ciiabs{}~$\lambda1334$ which are further analyzed in Section~\ref{sec:Measurements}. 

\subsection{Lya emitters and non-emitters} \label{sec:PANCAKEZ_LAEs_nonLAEs_composites}

As one of the intrinsically strongest emission lines in the Universe, \lya\ emission is commonly used to identify SFGs and to trace galaxy evolution \citep[e.g.,][]{ouchi20}. We observe a clear detection of Ly$\alpha$ emission in the full PANCAKEZ composite (see Figures~\ref{fig:Full-IndividualFilterComposites} and \ref{fig:Full-Composite_SpectrumAllFilter_NumberHistogram_AvgRedshifts}). We additionally divided the full PANCAKEZ sample into two sub-samples: \lya-emitters (LAEs) and non-LAEs. Traditionally, LAEs are often defined as galaxies with rest-frame \lya\ equivalent width $\mathrm{EW} \ge 20~\mathrm{\AA}$. However, due to the limited SNR, we adopted a more liberal definition classifying any galaxy with visibly detected Ly$\alpha$ emission as an LAE for our stacking purposes. The remaining galaxies are considered non-LAEs. From the total of $64$ galaxies, $12$ galaxies were determined to be LAEs ($10$ from JADES, $1$ from CEERS, $1$ from AURORA). The LAE vs. non-LAE classification is listed in Table~\ref{tab:PANCAKEZ_properties}. 

The composite spectra of the LAE and non-LAE samples are constructed in the same manner as described above. We display the G140M composite spectra for both samples in Figure~\ref{fig:restUV_LAE_nonLAE_comparisons} along with the full $64$ PANCAKEZ composite. The average $M_{\rm{UV}}$ and redshifts for the composites are $\left<M_{\rm{UV}}\right>_{\rm{LAEs}} = -20.3 \pm 0.8$, $\left<z\right>_{\rm{LAEs}} = 6.96 \pm 0.41$, $\left<M_{\rm{UV}}\right>_{\rm{non-LAEs}} = -19.6 \pm 1.3$, and $\left<z\right>_{\rm{non-LAEs}} = 6.89 \pm 0.65$, where the quoted errors are the standard deviation of the $M_{\rm{UV}}$ and redshift distributions. 

\section{Measurements and Analysis} \label{sec:Measurements}
To characterize the PANCAKEZ galaxies and to compare with galaxies in the post-EoR Universe at lower redshifts, we analyzed the composite spectra and extracted key properties such as absorption and emission line velocity centroid ($v_{\rm{cen}}$), equivalent width (EW), and full width at half maximum (FWHM). In this section we describe how these measurements were obtained from the PANCAKEZ stacks, as well as the referenced stacks of lower-z galaxies which we use for comparison in this work.

\subsection{LIS absorption} \label{sec:LIS_measurement}
\begin{figure}
\centering
\includegraphics[width=\columnwidth]{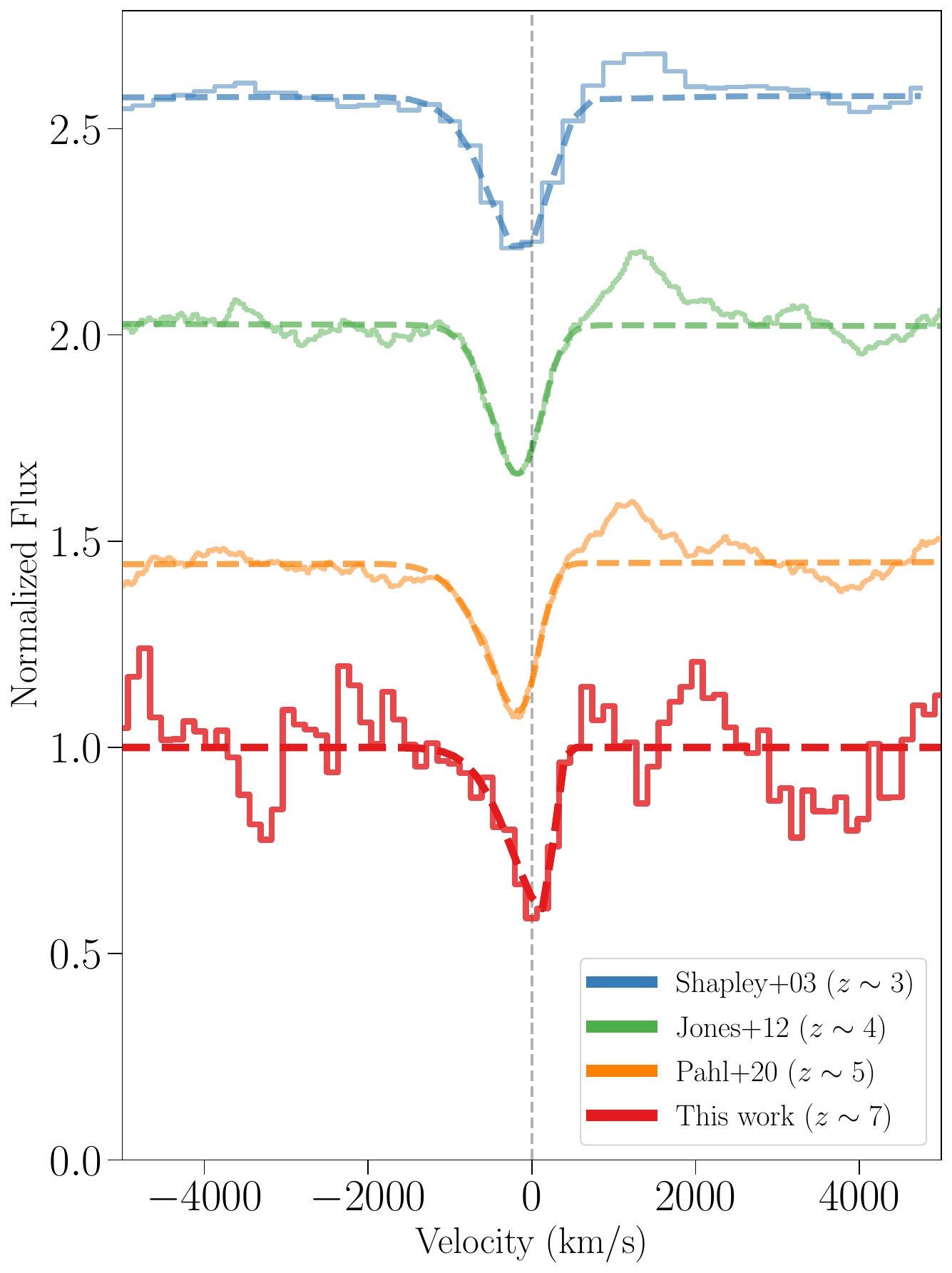}
\caption{Stacked LIS absorption profiles from \citet{Shapley2003_z3LBG} (blue), \citet{Jones2012_faintLBG} (green), \citet{Pahl2020_z5stacks} (orange) and this work (red). Solid lines show the respective mean LIS absorption and the dashed lines represent the best-fit skewed Gaussian model. Note that individual spectra are shifted vertically in increments of $0.5$ for display purposes. The averaged LIS absorption from our $z\sim7$ stack displays a smaller velocity offset (i.e., less blueshifted) along with overall weaker absorption strength compared to the lower-$z$ studies. 
\label{fig:MeanLISProfile}}
\end{figure}

The average LIS absorption line profile was derived by taking a mean of five strong LIS absorption features (\Siiiabs\ $\lambda1260$, \Oiabs\ $\lambda1302$, \Siiiabs\ $\lambda1304$, \Ciiabs\ $\lambda1334$, and \Siiiabs\ $\lambda1526$). We resample the composite spectrum onto a common velocity grid for each line, such that the rest-frame wavelength of the line corresponds to $v_{\rm{sys}}=0~\kms$, and then take a mean of the LIS lines to obtain the average profile. Figure~\ref{fig:MeanLISProfile} shows the averaged LIS absorption line profile of the full PANCAKEZ stack (red) along with other samples at lower redshifts: \citet{Shapley2003_z3LBG} at $z\sim3$, \citet{Jones2012_faintLBG} at $z\sim4$, and \citet{Pahl2020_z5stacks} at $z\sim5$. The lower-$z$ mean LIS profiles were made using the same methods as for the PANCAKEZ composite spectrum. Specifically, after re-normalizing the referenced stacks to the same wavelength range as PANCAKEZ ($1450 - 1500~\mathrm{\AA}$), the same LIS absorption features were isolated and averaged from respective composite spectra which resulted in the profiles shown in Fig~\ref{fig:MeanLISProfile}.
 
The averaged LIS absorption profiles were fit with two different models: (i) a symmetric Gaussian function and (ii) a skewed Gaussian function. The symmetric Gaussian is often used to determine the velocity centroid of the LIS absorption \citep[e.g.,][]{Du2018ApJ_specprops_LBGs} and is adopted for consistent comparisons with earlier work. However, as shown in Figure~\ref{fig:MeanLISProfile}, the blue side of the absorption profiles can be more extended than the red side due to galactic outflows \citep[e.g.,][]{Vasan2023}. Therefore, the skewed Gaussian is also adopted as a better model of the absorption line properties. For both models, velocity ranges corresponding to nearby fine structure emission lines (e.g., \ion{Si}{2}*~$\lambda1265$, $\lambda1309$, $\lambda1533$; \citealt{Jones2012_faintLBG,Shapley2003_z3LBG}) are masked out, and the fitting was conducted within the range $\pm5000 \: \rm km \;s^{-1}$ assuming a flat continuum. For the PANCAKEZ LAE stack where LIS absorption is not detected, we estimate $\rm{EW}_{\rm LIS}$ by taking the integrated flux over a fixed range ($-1000,600 \;\rm{kms}^{-1}$) divided out by the local continuum level (within $\pm 5000$~\kms). The associated $1\sigma$ uncertainty is derived by summing the error spectrum in quadrature over the same integrated range. The fixed velocity range is chosen such that it is comparable to the range encompassing the detected LIS absorption in the full PANCAKEZ stack. 

Best-fit LIS absorption properties for the PANCAKEZ stacks are recorded in Table \ref{tab:LISMeasurements} along with the lower-$z$ studies evaluated in this work. Values provided for the lower-$z$ composites were consistently measured using the same methods described above.

\subsection{Ly$\alpha$ emission}
\begin{figure}
\centering
\includegraphics[width=\columnwidth]{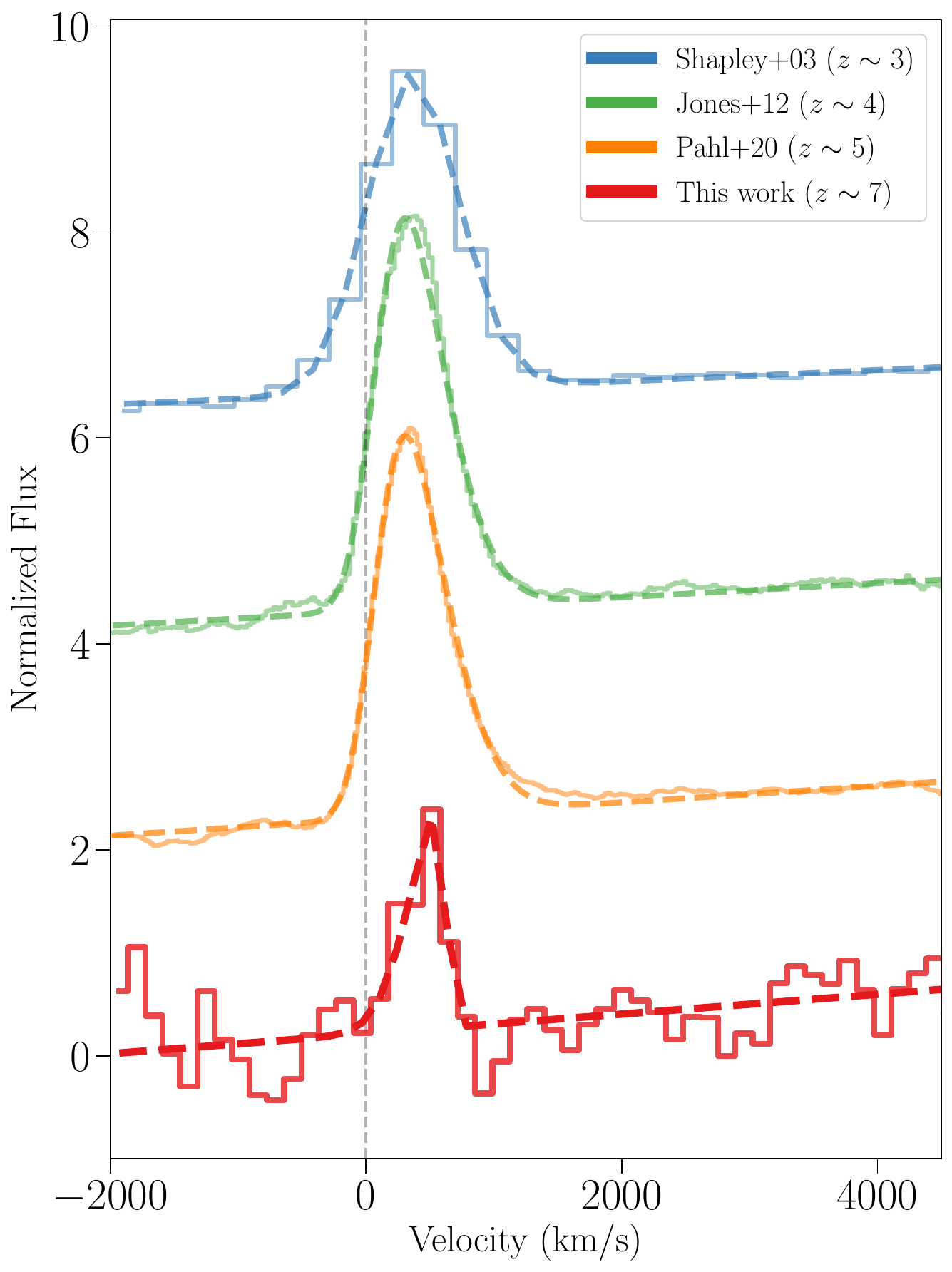}
\caption{Stacked Lya emission profiles from \citet{Shapley2003_z3LBG} (blue), \citet{Jones2012_faintLBG} (green), \citet{Pahl2020_z5stacks} (orange) and this work (red). Skewed Gaussian models are shown as a dashed line superimposed on the respective stack. A gray dashed line is drawn at the rest-frame wavelength of {\lya}. Individual spectra are shifted vertically in increments of $2$ for display purposes. The {\lya} emission from our $z\sim7$ stack is significantly weaker than in the lower-$z$ studies, which we attribute to suppression due to neutral \Hi\ in the IGM at $z\sim7$.
\label{fig:LyaProfiles}}
\end{figure}

Similar to the mean LIS absorption profile, we model the {\lya} emission using both the symmetric and skewed Gaussian profiles. Adapting to the more complex \lya\ feature, we use a modified fitting range of ($-2000, 5000  \: \rm{kms}^{-1}$) which aims to isolate the transmitted {\lya}, and excludes \ion{Si}{3} $\lambda1206$ absorption. We adopt a linear sloped continuum which provides an adequate fit to the observed continuum in this velocity range. In Figure~\ref{fig:LyaProfiles} we show the full PANCAKEZ stacked {\lya} profile (red) along with the profiles from lower-$z$ composite spectra studies \citep[$z\sim3$--5;][]{Shapley2003_z3LBG, Jones2012_faintLBG, Pahl2020_z5stacks}. {\lya} profiles for the lower-$z$ stacks were constructed in the same manner described above.

Given the intended scope of this work and the SNR of the continuum spectra, we do not correct for the IGM transmission before fitting the \lya\ spectral profile. To mitigate the effects on the measured {\lya} EW, we follow the approach by \citet{Kornei2010_LBG_stellarpops} and measure an additional continuum level ($1225 - 1255 \ \mathrm{\AA}$; $c_{\rm{red}}$) for each spectrum. The {\lya} emission EW is then calculated by dividing the enclosed flux (modeled flux subtracted by the local continuum) by $c_{\rm{red}}$. For the PANCAKEZ non-LAE stack, we estimate the $\rm{EW}_{\rm Ly\alpha}$ by integrating the flux over a fixed velocity range ($-500,1500 \; \rm{kms}^{-1}$). The associated $1\sigma$ uncertainty is derived from the error spectrum by summing in quadrature over the same integrated range. The fixed velocity range is determined such that it encompasses the same range to that of the \Lya\ emission detected in the full PANCAKEZ stack. Resulting values for the \Lya\ emission line properties are recorded in Table \ref{tab:LyaMeasurements}. Values are also provided for the lower-$z$ composite stacks which were measured in the same manner as described for the PANCAKEZ stack.

\subsection{Nebular emission lines} \label{sec:neb_emission_measurements}
We analyze the nebular emission lines in the composite spectra to assess possible systematic error sources and to characterize the physical conditions of the galaxies in our sample. Each nebular emission line of interest in the composite spectra was fitted with a single Gaussian profile, yielding residual fluxes $<5\%$ of the total integrated flux. For \Oiii $\lambda\lambda4960,5008$, the measured fluxes agree with the expected theoretical ratio of 2.98 within $\sim 5\%$. We thus adopt a $5\%$ systematic uncertainty for all flux ratios in this work.

We additionally assess the accuracy of the spectroscopic redshifts in the composite spectra (discussed in Section~\ref{sec:spectroscopic_redshifts}) using the velocity centroids ($v_{\rm{cen}}$) of the nebular emission lines. In Table~\ref{tab:nebEmissionLines} we report the $v_{\rm{cen}}$ of \Hb\ and \Oiii\ for each stacked PANCAKEZ spectrum. Across all three stacks, we find the average nebular $\left< v_{\rm{cen}} \right> = -0.7 \ \kms$ with a standard deviation of 6.8~\kms.
This represents an upper limit on the true uncertainty in the velocity centroid of individual spectral features. This uncertainty is similar to those measured at $z \sim 2$--3 from ground-based telescopes \citep[e.g., ][]{steidel14, kriek15}, and validates that redshift uncertainties do not significantly impact the results of our analysis.

Furthermore, we estimate the effective spectral resolution of our composite spectra using the full width at half maximum (FWHM) of the nebular emission lines (listed in Table~\ref{tab:nebEmissionLines}). With NIRSpec achieving $R\sim 1000$ in the medium resolution grating, this sets an instrumental limit of $\rm{FWHM}\simeq 300 \; \mathrm{km~s}^{-1}$. The average nebular FWHM across all stacks is 343~\kms\ with a standard deviation of 14~\kms. Assuming an intrinsic FWHM of $\sim 240~\mathrm{km~s}^{-1}$ \citep{steidel14}, the PANCAKEZ composite achieves a comparable effective resolution to that of the NIRSpec instrument. This further confirms that the systemic redshifts used in this work are accurate. Notably, we can conclude that the LIS absorption and \Lya\ emission lines with FWHM~$\gtrsim 500$~\kms\ in the composite spectra (Tables~\ref{tab:LISMeasurements}, \ref{tab:LyaMeasurements}) are significantly broader than the effective spectral resolution, indicating that these widths reflect intrinsic velocity structure in the line profiles.

\begin{deluxetable}{llccc}
\setlength{\tabcolsep}{2pt} 
\tabletypesize{\scriptsize}
\tablewidth{\columnwidth}
\tablecaption{LIS Measurements \label{tab:LISMeasurements}}
\tablehead{
\colhead{Study} & \colhead{} & \colhead{Velocity Centroid} & \colhead{FWHM} & \colhead{EW} \\
\colhead{} & \colhead{} & \colhead{($\rm km~s^{-1}$)} & \colhead{($\rm km~s^{-1}$)} & \colhead{(\AA)}
}
\startdata 
(\textit{Skew Gaussian}) &  & &  &  \\ 
\vspace{2mm}
{Shapley+03 ($z\sim3$)} & & $-220.2^{+17.6}_{-16.5}$ & $871^{+28}_{-23}$ & $-1.82^{+0.05}_{-0.05}$ \\
\vspace{2mm}
{Jones+12 ($z\sim4$)} & & $-217.4^{+7.7}_{-7.1}$ & $718^{+13}_{-11}$ &$-1.48^{+0.02}_{-0.02}$ \\
\vspace{2mm}
{Pahl+20 ($z\sim5$)} & & $-332.7^{+6.9}_{-8.4}$ & $763^{+13}_{-12}$ & $-1.52^{+0.02}_{-0.02}$ \\
\vspace{0.8mm} 
{     } & Full Stack& $-132.2 ^{+85.6}_{-108.0}$ & $589^{+216}_{-125}$ & $-1.18^{+0.26}_{-0.28}$ \\
\vspace{0.8mm} 
{This Work ($z\sim7$)} &  LAEs & - & - & $-0.27^{+0.42}_{-0.42}$ \\
{     } & Non-LAEs & $-110.6^{+77.3}_{-99.1}$ & $648^{+146}_{-161}$ & $-1.36^{+0.21}_{-0.41}$ \\
\hline
(\textit{Normal Gaussian}) &  & &  & \\ 
\vspace{2mm}
{Shapley+03 ($z\sim3$)} & & $-177.3^{+18.1}_{-18.1}$ & $869^{+45}_{-45}$ & $-1.78^{+0.08}_{-0.08}$ \\ 
\vspace{2mm}
{Jones+12 ($z\sim4$)} & & $-197.3^{+7.9}_{-7.9}$ & $722^{+19}_{-19}$ &$-1.47^{+0.04}_{-0.04}$ \\
\vspace{2mm}
{Pahl+20 ($z\sim5$)} & & $-255.0^{+8.6}_{-8.6}$ & $733^{+21}_{-21}$ & $-1.44^{+0.04}_{-0.04}$ \\
\vspace{0.8mm}
{     } & Full Stack & $-23.5 ^{+50.7}_{-50.7}$ & $583^{+123}_{-123}$ & $-1.14^{+0.21}_{-0.21}$ \\
\vspace{0.8mm}
{This Work ($z\sim7$)} & LAEs & - & - & $-0.27^{+0.42}_{-0.42}$ \\
{     } & Non-LAEs & $-59.1 ^{+61.1}_{-61.1}$ & $684^{+149}_{-149}$ & $-1.35^{+0.26}_{-0.26}$
\enddata
\tablecomments{LIS measurements from skewed and normal Gaussian best fits. As a reminder, we adopt the convention of negative EW values to represent absorption. Literature best-fit values were remeasured from their respective composites in the same manner as this work.}
\end{deluxetable}

\begin{deluxetable}{llccc}
\setlength{\tabcolsep}{2pt} 
\tabletypesize{\scriptsize}
\tablecaption{Ly$\alpha$ Measurements \label{tab:LyaMeasurements}}
\tablewidth{\columnwidth}
\tablehead{
\colhead{Study} & \colhead{} & \colhead{Velocity Centroid} & \colhead{FWHM} & \colhead{EW} \\
\colhead{} & \colhead{} & \colhead{($\rm km~s^{-1}$)} & \colhead{($\rm km~s^{-1}$)} & \colhead{($\AA$)}
}
\startdata 
(\textit{Skew Gaussian}) & & & & \\ 
\vspace{2mm}
{Shapley+03} & & $377.5^{+3.9}_{-5.3}$ & $834^{+8}_{-7}$ & $15.44^{+0.12}_{-0.13}$ \\ 
\vspace{2mm}
{Jones+12} & & $393.4^{+4.1}_{-4.2}$ & $645^{+9}_{-8}$ & $16.21^{+0.36}_{-0.28}$ \\
\vspace{2mm}
{Pahl+20} & & $405.8^{+1.7}_{-1.3}$ & $652^{+2}_{-3}$ & $15.48^{+0.06}_{-0.05}$ \\
\vspace{0.7mm}
{     } & Full Stack & $404.7 ^{+82.0}_{-69.8}$ & $361^{+131}_{-112}$ & $4.60^{+0.95}_{-1.30}$ \\
\vspace{0.7mm}
{This work} & LAEs & $292.9^{+18.8}_{-17.1}$ & $565^{+34}_{-40}$ & $40.47^{+2.84}_{-3.61}$ \\
{     } & Non-LAEs & - & - & $1.99^{+1.78}_{-1.78}$\\
\hline
(\textit{Normal Gaussian}) & & & & \\ 
\vspace{2mm}
{Shapley+03} & & $367.0^{+4.1}_{-4.1}$ & $889^{+11}_{-11}$ & $15.54^{+0.182}_{-0.182}$ \\ 
\vspace{2mm}
{Jones+12} & & $361.2^{+2.2}_{-2.2}$ & $651^{+5}_{-5}$ &$16.41^{+0.129}_{-0.129}$ \\
\vspace{2mm}
{Pahl+20} & & $356.8^{+2.8}_{-2.8}$ & $649^{+7}_{-7}$ & $15.05^{+0.158}_{-0.158}$ \\
\vspace{0.7mm}
{     } & Full Stack & $442.1 ^{+28.5}_{-28.5}$ & $411^{+70}_{-70}$ & $4.51^{+0.70}_{-0.70}$ \\
\vspace{0.7mm}
{This work} & LAEs & $283.8 ^{+7.3}_{-7.3}$ & $578^{+18}_{-18}$ & $40.7^{+1.22}_{-1.22}$ \\
{     } & Non-LAEs & - & -  & $1.99^{+1.78}_{-1.78}$ \\
\enddata
\tablecomments{Ly$\alpha$ measurements from skewed and normal Gaussian best fit models. As a reminder, we adopt the convention of positive EW values to represent emission. Literature best-fit values were remeasured from their respective composites in the same manner as this work. We adopt the method of \citet{Kornei2010_LBG_stellarpops} to measure {\lya} EW values.}
\end{deluxetable}

\begin{deluxetable}{llccc}
\setlength{\tabcolsep}{2pt} 
\tabletypesize{\scriptsize}
\tablewidth{\columnwidth}
\tablecaption{Nebular Emission Line Measurements \label{tab:nebEmissionLines}}
\tablehead{
\colhead{} & \colhead{} & \colhead{Velocity Centroid} & \colhead{FWHM} & \colhead{EW} \\
\colhead{} & \colhead{} & \colhead{($\rm km~s^{-1}$)} & \colhead{($\rm km~s^{-1}$)} & \colhead{(\AA)}
}
\startdata 
\hline
{     } & Full stack & $-3.22 \pm 2.12$ & $356 \pm 5$ & $98.68 \pm 1.34$ \\
{H$\beta_{\lambda4862}$} & LAEs & $-2.27 \pm 3.39$ & $337 \pm 8$ & $172.52 \pm 3.93$ \\
{     } & non-LAEs & $-6.45 \pm 2.48$ & $364\pm 6$ & $74.09 \pm1.15$ \\
\hline
{     } & Full stack & $4.43 \pm 1.72$ & $316 \pm 4$ & $205.35 \pm 2.51$ \\
{$\rm{[OIII]}_{\lambda4960}$} & LAEs & $14.62 \pm 2.01$ & $328 \pm 5$ &$379.16 \pm 5.28$ \\
{     } & non-LAEs & $-2.38 \pm 1.91$ & $351\pm 5$ & $168.94 \pm 2.10$ \\
\hline
{     } & Full stack & $-5.32 \pm 0.88$ & $342 \pm 2$ & $553.20 \pm 3.24$ \\
{$\rm{[OIII]}_{\lambda5008}$} & LAEs & $1.62 \pm0.77$ & $343 \pm 2$ &$744.20 \pm3.80$ \\
{} & non-LAEs & $-6.91 \pm 1.24$ & $351 \pm 3$ & $476.10 \pm 3.84$ \\
\hline
\enddata
\tablecomments{Nebular emission line properties measured from normal Gaussian fits. As a reminder, we adopt the convention of negative EW values for absorption and positive values for emission.}
\end{deluxetable}

\section{Discussion}
\label{sec:discussion}

\subsection{A representative sample at $z>6$}
\label{sec:representative_sample}
The underlying goal of the PANCAKEZ study is to investigate the rest-UV spectroscopic properties of galaxies found in the EoR. As mentioned previously, reionization is not thought to be dominated by bright, massive galaxies. Instead, fainter, less massive, more numerous galaxies are thought to provide the bulk of the ionizing photon budget \citep[e.g.,][]{Robertson2010_ionizationEQ,Bouwens2022_Lfunction_z2to9, Borsani2024ApJ_R100_stacks}. These fainter EoR galaxies, however, are challenging to observe individually in the rest-UV requiring very deep spectra with long integration times. 
The PANCAKEZ study overcomes this challenge by taking a statistical average (i.e., stack) across a range of selected galaxies which ideally encompasses the dominant EoR population. To that end, it is important to investigate (i) how representative the PANCAKEZ sample is with respect to the broader EoR galaxy population, and (ii) any inherent selection effects in the sample. 

\subsubsection{Nebular emission line diagnostics} \label{sec:neblinediagonstic_O32_R23}
\begin{figure}
\centering\includegraphics[width=0.95\columnwidth]{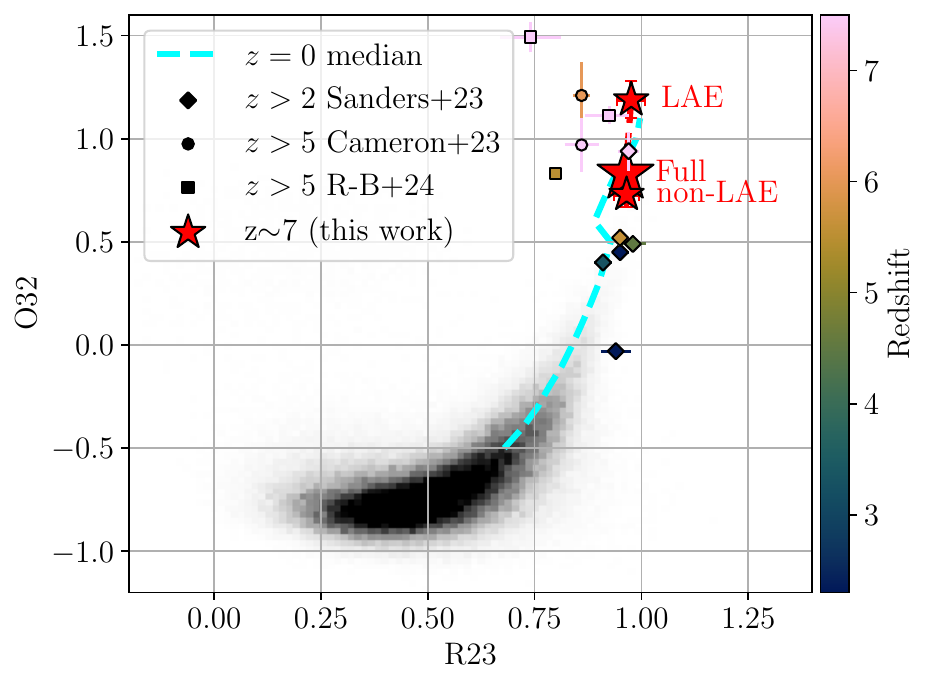}
\caption{The reddening-corrected O32 vs. R23 diagram for the PANCAKEZ sample (red stars), compared with low-redshift star-forming galaxies from the Sloan Digital Sky Survey (SDSS; \citealt{tremonti04}) shown in grayscale. The cyan dashed line marks the median ridgeline of the SDSS sample within the high-ionization/excitation tail. Also shown are recent JWST composites from \citet{Cameron2023_ISM_z5p5, Sanders23_ionizationpropz2to9, Borsani2024ApJ_R100_stacks}  (color-coded by redshift). The PANCAKEZ sample, with its high ionization/excitation, lies on the extreme tail of the low-$z$ galaxy distribution and is representative of typical galaxies at $z > 6$.
\label{fig:O32_R23}}
\end{figure}

\begin{deluxetable*}{llccc}
\tablewidth{0pt} 
\tablecaption{Strong-line ratios of the PANCAKEZ sample \label{tab:strong-line-ratios}}
\tablehead{
\colhead{Name} & \colhead{Definition} & \multicolumn{3}{c}{Value} \\
\colhead{} & \colhead{} & \colhead{Full} & \colhead{LAE} & \colhead{non-LAE}}
\startdata 
O3 & $\log$ [\ion{O}{3}]~$\lambda5007/\mathrm{H}\beta$ & $0.789 \pm 0.065$ & $0.838 \pm 0.069$ & $0.784 \pm 0.066$ \\ 
O2 & $\log$ [\ion{O}{2}]~$\lambda\lambda3726,29/\mathrm{H}\beta$ & $-0.041 \pm 0.082$ & $-0.351 \pm 0.11$ & $0.055 \pm 0.086$  \\
R23 & $\log$ ([\ion{O}{3}]~$\lambda\lambda4960,5008$ + [\ion{O}{2}]~$\lambda\lambda$3726,29$)/\mathrm{H}\beta$ & $0.963 \pm 0.056$ & $0.976 \pm 0.061$ & $0.965 \pm 0.058$  \\
O32 & $\log$ [\ion{O}{3}]~$\lambda5007$ / [\ion{O}{2}]~$\lambda\lambda$3726,29 & $0.830 \pm 0.082$ & $1.19 \pm 0.11$ & $0.730 \pm 0.085$ \\
Ne3O2 & $\log$ [\ion{Ne}{3}]~$\lambda3869$ / [\ion{O}{2}]~$\lambda\lambda$3726,29 & $-0.37 \pm 0.29$ & $-0.11 \pm 0.14$ & $-0.44 \pm 0.46$ \\
\enddata
\end{deluxetable*}

Using the flux ratios of the nebular emission lines from Section~\ref{sec:neb_emission_measurements}, we examine widely-used nebular diagnostics to investigate typical physical conditions of the PANCAKEZ galaxies and compare them with other high-redshift samples. For each stack, the flux ratios are determined from the weighted average composite spectra which incorporate the G140M, G235M, and G395M gratings. This approach ensures comprehensive spectral coverage while minimizing sample variations across different wavelengths. Dust attenuation is derived from the H$\gamma$/H$\beta$ ratio using the \citet{cardelli89} extinction curve with $R_V=3.1$, assuming an intrinsic H$\gamma$/H$\beta = 0.468$. The choice of extinction curve does not significantly affect the results for the line ratios used herein. 
For all composite spectra, we adopt a single dust attenuation correction $E(B-V) = 0.06 \pm 0.16$ obtained from the full stack. This assumption is necessary for the LAE sample in which the H$\gamma$ emission is too faint for robust measurement. Reddening-corrected measurements of various commonly used strong-line ratios are summarized in Table~\ref{tab:strong-line-ratios} for the full sample, LAE, and non-LAE composites.

Figure~\ref{fig:O32_R23} shows the O32 vs. R23 diagram (see Table~\ref{tab:strong-line-ratios} for definitions), which approximately traces the ionization state and metallicity of the ISM nebular gas. As with other $z > 6$ galaxies, the PANCAKEZ sample occupies the high O32 and R23 tail of the local SFG distribution, corresponding to high ionization parameter and low metallicity \citep[e.g.,][]{tremonti04}. 

The PANCAKEZ stacked spectra are within the region occupied by the $z = 6.5$--9.3 sample reported by \citet{Sanders23_ionizationpropz2to9}, but are slightly offset compared to $z>5$ samples reported by \citet{Cameron2023_ISM_z5p5} and  \citet{Borsani2024ApJ_R100_stacks}, suggesting that our targets may be slightly higher in metallicity. While the full and non-LAE samples exhibit comparable ionization and excitation levels, the LAE sample displays significantly higher ionization (O32) which exceeds that of typical galaxies at $z\sim 7$. Overall, the PANCAKEZ sample appears to be representative of typical galaxies at $z > 6$ while the LAE sample corresponds to the high-ionization tail of the population. 

We can estimate the typical gas metallicity (e.g., gas-phase oxygen abundance) in our sample from the strong-line ratios presented in Table~\ref{tab:strong-line-ratios}. 
We estimate metallicity using the direct-method calibrations from \citet{sanders24a}. The O2 and O32 ratios for our sample provide the strongest constraints and suggest gas-phase abundances $12 + \log(\rm{O}/\rm{H}) \sim 7.9$ (for the full sample composite), 7.6 (LAE), and 8.0 (non-LAE). The uncertainty in O/H abundances is $\sim$0.2 dex based on scatter in these calibrations. O3 and R23 ratios for our sample lie on the flat portion (``turnover'') in their relation with $12 + \log(\rm{O}/\rm{H})$; they are consistent with the values noted above but offer limited constraining power. Ne3O2 is likewise consistent but has the highest measurement uncertainty. 

We also measure O/H abundance using the standard direct ``$\mathrm{T_e}$ method.'' Since the \Oiii~$\lambda4363$ emission is not well detected in the subsamples and its spectral profile varies across the stacks due to small sample statistics, we report the electron temperature ($T_e = 21000 \pm 2000~\mathrm{K}$) only for the full stack. 
We then obtain the total O abundance by combining the $O^+/H^+$ and $O^{++}/ H+$ as $12 + \log(O/H) = 7.63 \pm 0.03$ (statistical) $\pm 0.03$ (systematic) \footnote{The systematic uncertainty is constributed by the assumed $T_e(O^+) = 0.7 T_e(O^{++}) + 3000~\mathrm{K}$ \citep{esteban09} and $n_e = 100~\mathrm{cm}^{-3}$.}. This O/H metallicity is $\sim 0.3$ dex lower than the estimate from the strong-line method.

Both the $T_e$ and gas metallicity values are similar to those of other star-forming galaxies at $z>6$ measured with both the direct $\mathrm{T_e}$ method and via strong-line calibrations \citep[e.g.,][]{bunker23, Sanders23_ionizationpropz2to9, curti23, morishita24}, and consistent with the stacking analysis of nebular emission at similar redshift by \cite{Hu2024_ISM_stacks}. The LAE stack has lower metallicity by $\sim$0.3 dex, which is comparable to the measurement uncertainty, but supports a picture whereby LAEs in our sample tend to be younger or less evolved. 

\subsubsection{UV luminosity} \label{MUV_at_z6}

The spectroscopic properties of interest in this work have been shown to correlate with luminosity (e.g., absolute rest-frame UV magnitude $\rm{M}_{\rm{UV}}$), where fainter galaxies tend to have stronger \Lya\ emission and weaker LIS absorption \citep[e.g.,][]{Du2018ApJ_specprops_LBGs,Jones2012_faintLBG}. We thus examine the $\rm{M}_{\rm{UV}}$ distribution of the PANCAKEZ galaxies in relation to typical SFGs in the epoch of reionization. 

A recent study at $z\geq5$ by \citet{Borsani2024ApJ_R100_stacks} shows that the vast majority of their sample of $>400$ SFGs have $\rm{M}_{\rm{UV}} \gtrsim -20$. Their sample has an average $\rm{M}_{\rm{UV}} \approx -18.6 \pm 0.1$ in the redshift interval $6<z<7$, whereas the $42$ PANCAKEZ sources within the same redshift interval are typically one magnitude brighter ($\left< \rm{M}_{\rm{UV}} \right> = -19.6$). We note, however, that only $4/42$ PANCAKEZ galaxies exceed the brightest source reported by \citet{Borsani2024ApJ_R100_stacks} in this redshift range ($6<z<7$). A similar difference persists at higher redshift $7 < z  < 8$, with average $\rm{M}_{\rm{UV}}  \approx -19.0 \pm 0.1$ from \citet{Borsani2024ApJ_R100_stacks}, and $\left< \rm{M}_{\rm{UV}} \right> = -19.8$ for the $18$ PANCAKEZ sources in this range. The stacking sample of \cite{Kumari2024_LAEstacks} is of intermediate luminosity, with average $\rm{M}_{\rm{UV}} \approx -19.0$ for their $z>6.3$ sources. We thus find that the PANCAKEZ sample probes $\sim0.5$--1 magnitude higher average luminosity compared to previous composite samples of \citet{Borsani2024ApJ_R100_stacks} and \citet{Kumari2024_LAEstacks}. 

Overall, the luminosity distribution of galaxies at this epoch is reasonably described by a Schechter function. \cite{Bouwens2021_LF_z29} measure the characteristic UV absolute magnitude as $M^{*}_{\rm{UV}}= -21.15 \pm0.13$ at $z\sim7$. The PANCAKEZ sample ($\left<\rm{M}_{\rm{UV}}\right> = -19.7$ with a sample standard deviation of 1.1) is typically $\sim$1.5 magnitudes fainter than $M^{*}$, spanning a range of $-18 \lesssim M_{\rm{UV}} \lesssim -22$ which corresponds to $0.1 \lesssim L^{*} \lesssim 2$. While PANCAKEZ contains on average brighter objects than previous composite spectra analyses, the overall spread of our galaxies lies reasonably within the boundaries of what has been seen of typical EoR galaxies when compared to blank-field sources. In summary, the PANCAKEZ sample is representative of the moderately bright end($\gtrsim 0.1 L^*$) of the UV luminosity function at $z>6$.

We finally comment on the galaxy $M_{\rm{UV}}$ distributions from the lower-$z$ stacks which we draw direct comparisons to for this work. The stacks by \citet[][$\left<\rm{M}_{\rm{UV}}\right> \approx -20.91$ at $z\sim4$]{Jones2012_faintLBG} and \citet[][$\left<\rm{M}_{\rm{UV}}\right> \approx -21.33$ at $z\sim5$]{Pahl2020_z5stacks} probe overall $\sim3$--4 times higher UV luminosity than PANCAKEZ. We refer readers to, e.g., Figure~5 of \citealt{Pahl2020_z5stacks} for a distribution of the $z\sim2$--5 samples which we compare to in this work. We note that at fixed redshift, less luminous (and less massive, smaller, and bluer) galaxies tend to have stronger \Lya\ emission and weaker LIS absorption \citep[e.g.,][]{Du2018ApJ_specprops_LBGs}. Some of the redshift evolution which we explore in later sections is likely due to scaling relations given the lower UV luminosity of our $z\sim7$ sample.

\subsubsection{Selection bias}  
In Section~\ref{sec:stacking_PANCAKEZ} we discussed the sample selection for this work and its rationale. We now discuss potential selection biases. 

First, our requirement of precise spectroscopic redshifts caused us to reject $92$ sources which lacked detectable nebular emission lines in medium-resolution spectra. Galaxies that are still UV-bright but have recently stopped forming stars ($\gtrsim 10 \rm{Myrs}$ ago, i.e. post-starbursts) would not exhibit strong nebular emission lines and thus be removed from the fiducial sample. An example is the $z=7.3$ galaxy RUBIES-UDS-QG-z7 studied by \cite{Valentino2025}. Our sample thus represents only the actively star-forming population at $z>6$. However, only 10 of the rejected galaxies with insufficient emission line redshifts would have passed our other quality cuts (e.g., continuum SNR), such that our final sample contains $\gtrsim$85\% of the total observed UV-bright population.

Our second contributing selection effect is the continuum SNR criterion. Limiting the sample in this way ensures a suitable SNR in the composite spectrum but biases our result towards galaxies with higher continuum SNR. These sources are inherently brighter, or in some cases have longer-than-average integration times. 
The result is that the PANCAKEZ sample spans the moderately bright end of the UV luminosity function, centered around absolute $\rm{M}_{\rm{UV}} \sim -20$ (Figure~\ref{fig:M_UV_histogram_64gals}), $\sim 1.5 $ magnitudes fainter than $M^{*}_{UV}= -21.15$ reported in \citet{Bouwens2022_LF}. 

\subsection{LIS absorption at $z\geq6$}

\begin{figure*}
\centering
\includegraphics[width=\textwidth]{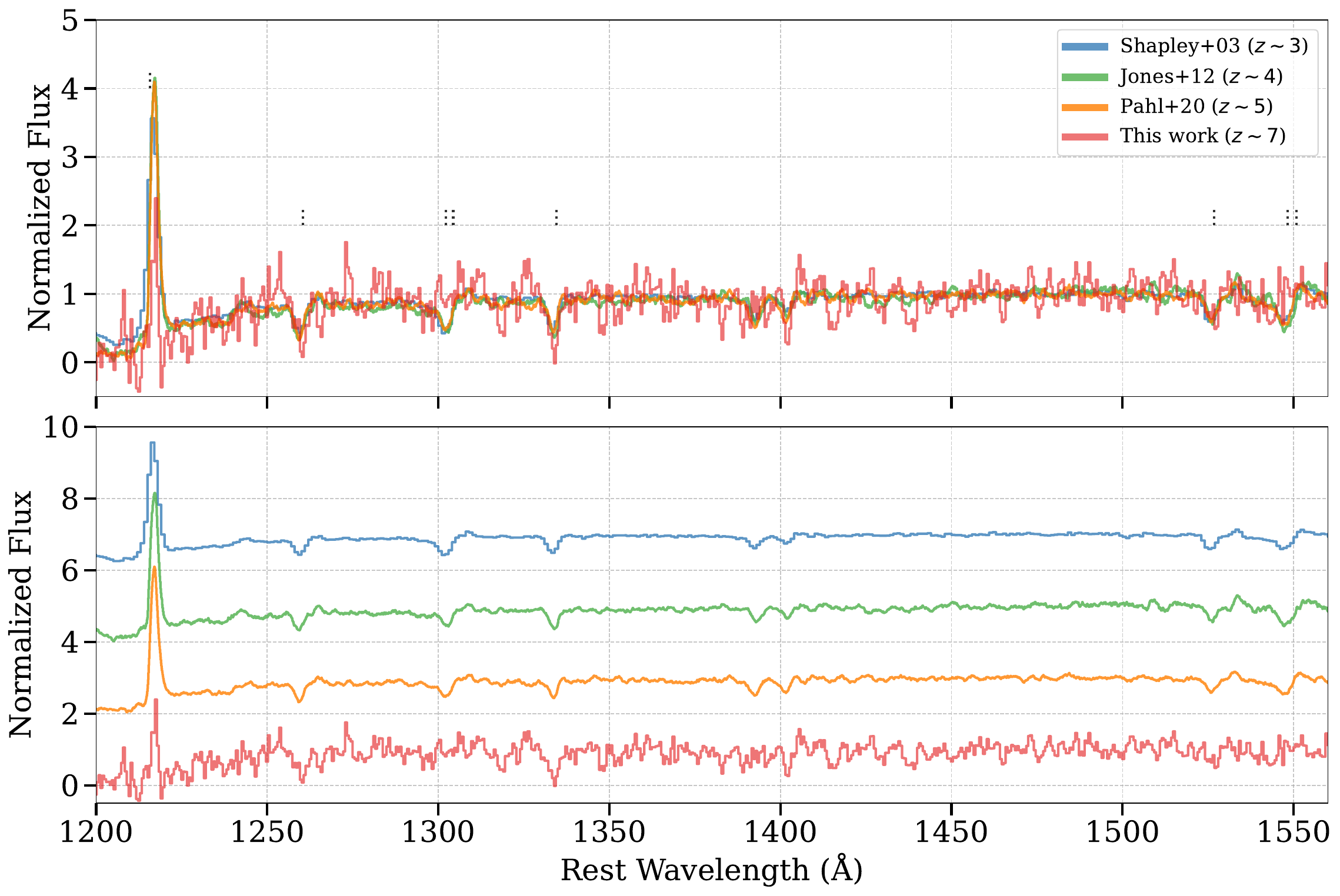}
\caption{The rest-UV composite spectra from \citet[][blue]{Shapley2003_z3LBG}, \citet[][green]{Jones2012_faintLBG}, and \citet[][orange]{Pahl2020_z5stacks} overlaid on our full $64$ galaxy PANCAKEZ composite (red). All spectra are consistently normalized at rest-frame 1450--1500~\AA. Black dotted lines in the upper panel mark various spectral features of interest: \Hi$\lambda1216$ (\Lya), \Siiiabs$\lambda1260$, \Oiabs$\lambda 1302$, \Siiiabs$\lambda 1304$, \Ciiabs$\lambda 1334$, \Siiiabs$\lambda 1526$, and \Civ$\lambda \lambda 1548,1550$. Overall the PANCAKEZ composite shows a continuum shape and spectral features similar to lower-$z$ studies, although LIS absorption lines and \Lya\ emission are weaker in the PANCAKEZ $z\sim7$ composite.
}
\label{fig:restUV_literature_comparisons}
\end{figure*}

A principal goal for this study is to measure LIS absorption lines in EoR galaxies. 
In particular, LIS lines can probe the neutral gas covering fraction which in turn regulates the escape of ionizing radiation \citep[e.g.,][]{Reddy2016ApJ_gascovfrac}. In this section we discuss the PANCAKEZ results, compare them with equivalent lower-$z$ studies (e.g., see Fig \ref{fig:restUV_literature_comparisons}), and discuss the implications for EoR galaxies. 
For consistency across compared studies, all composite spectra from the literature are re-normalized with the method described in Section~\ref{sec:stacking_PANCAKEZ}. We have additionally re-measured LIS and Ly$\alpha$ line properties following the same methods as described in Section~\ref{sec:Measurements}. As a reminder, re-measured literature values are provided alongside PANCAKEZ results in Tables~\ref{tab:LISMeasurements} and \ref{tab:LyaMeasurements}. 

\subsubsection{The relation between LIS and Ly$\alpha$}
\label{sec:LIS_Lya}

\begin{figure}[ht!]
\centering
\includegraphics[width=\columnwidth]{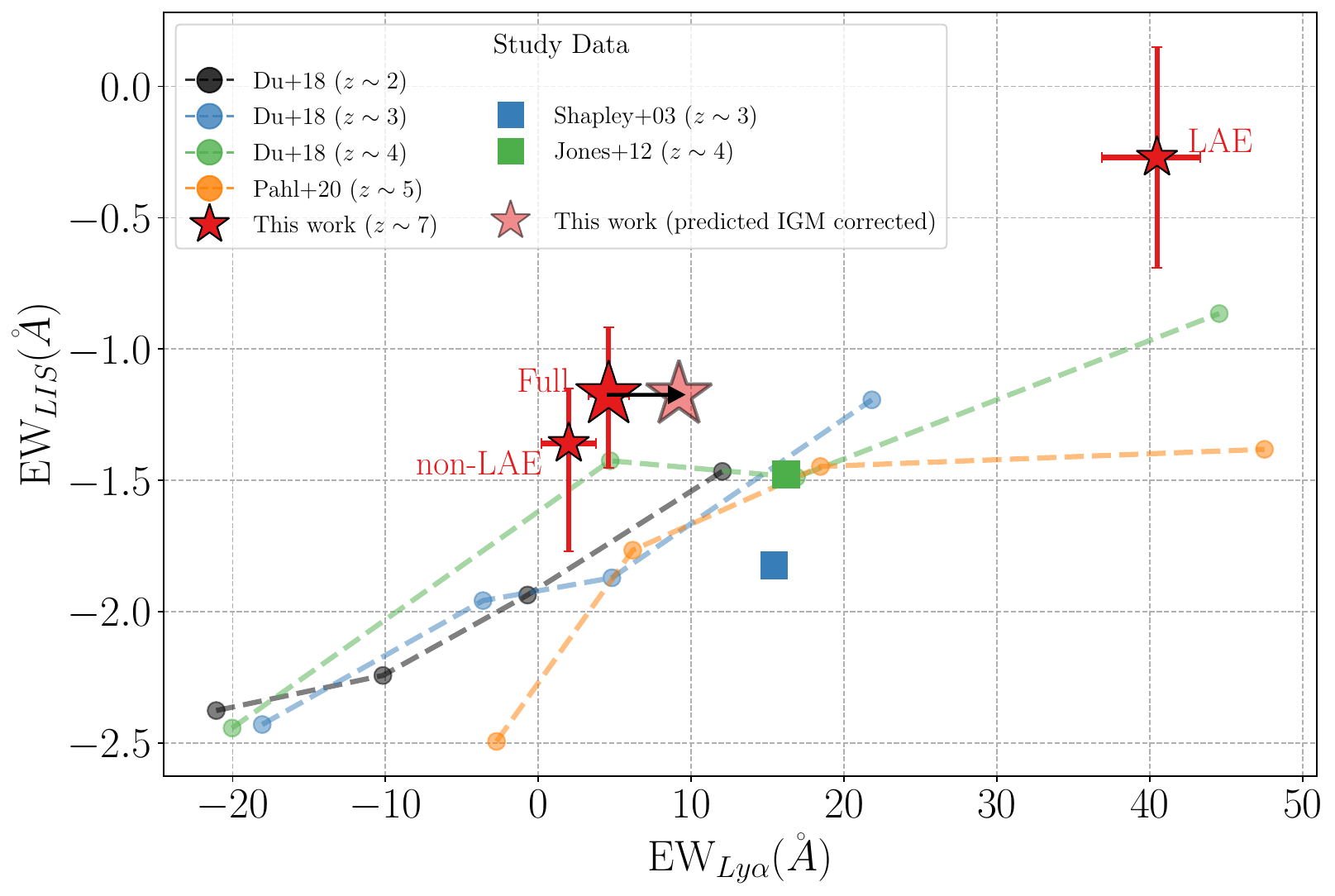}
\caption{$\rm{EW}_{\rm{LIS}}$ vs. $\rm{EW}_{\rm{Ly}\alpha}$. Dashed lines with circle markers represent the transmitted (uncorrected for IGM attenuation of \Lya) values reported in \citet{Pahl2020_z5stacks, Du2018ApJ_specprops_LBGs}. These data show a clear correlation with weaker LIS absorption corresponding to stronger \Lya\ emission. Square and star markers represent values measured from our skewed Gaussian fits. Symbol colors correspond to the redshift of each sample: black ($z\sim2$), blue ($z\sim3$), green ($z\sim4$), orange ($z\sim5$), and red ($z\sim7$). The black arrow shows the effect of correcting for $\sim$50\% IGM attenuation of the \Lya\ emission as found by \citet{Tang2024_Lya_SFGs_z613}. We note that for this plot and other figures, negative/positive EW values correspond to absorption/emission. The full PANCAKEZ composite shows clear deviation from the lower-$z$ trends with overall weaker LIS absorption and suppressed {\lya} emission. 
}
\label{fig:LIS_EW_vs_LYA_EW_UPDATED}
\end{figure}

Several low-$z$ studies have established the anti-correlation between $\rm{EW}_{\rm{LIS}}$ and $\rm{EW}_{\rm{Ly}\alpha}$: galaxies with stronger Ly$\alpha$ emission have weaker LIS absorption on average \citep[e.g.][]{Shapley2003_z3LBG,Vanzella2009LBGs_z456, Jones2012_faintLBG,Faisst2016_metallicity_z5,  Du2018ApJ_specprops_LBGs, Pahl2020_z5stacks}. Conversely, the presence of stronger LIS absorption correlates with weaker Ly$\alpha$ emission (and stronger \Lya\ absorption). 
This relationship is explained by a physical picture whereby LIS ions are found predominantly in neutral \Hi\ gas, as expected given their ionization potentials. The same gas which produces LIS absorption is thus responsible for scattering \Lya\ and regulating its escape fraction. Moreover, the \Lya\ emission escape fraction has been shown to positively correlate with LyC {\fesc} \citep[e.g.,][]{Steidel2018_KLCS_z3, Pahl2021_LyC_z3, Flury22_LzLCS_I}.
A direct way to facilitate a higher LyC {\fesc} is to have a low HI {\fcov}, which would be evident through weaker LIS absorption lines. We plot $\rm{EW}_{\rm LIS}$ versus $\rm{EW}_{\rm Ly\alpha}$ in Figure \ref{fig:LIS_EW_vs_LYA_EW_UPDATED} for the PANCAKEZ stacks, along with previous stacking results at redshifts $z\sim2$--5. This figure has been adapted from \citet{Pahl2020_z5stacks} to show transmitted values (not corrected for IGM attenuation) of {\lya}. 

\begin{figure}
\centering
\includegraphics[width=\columnwidth]{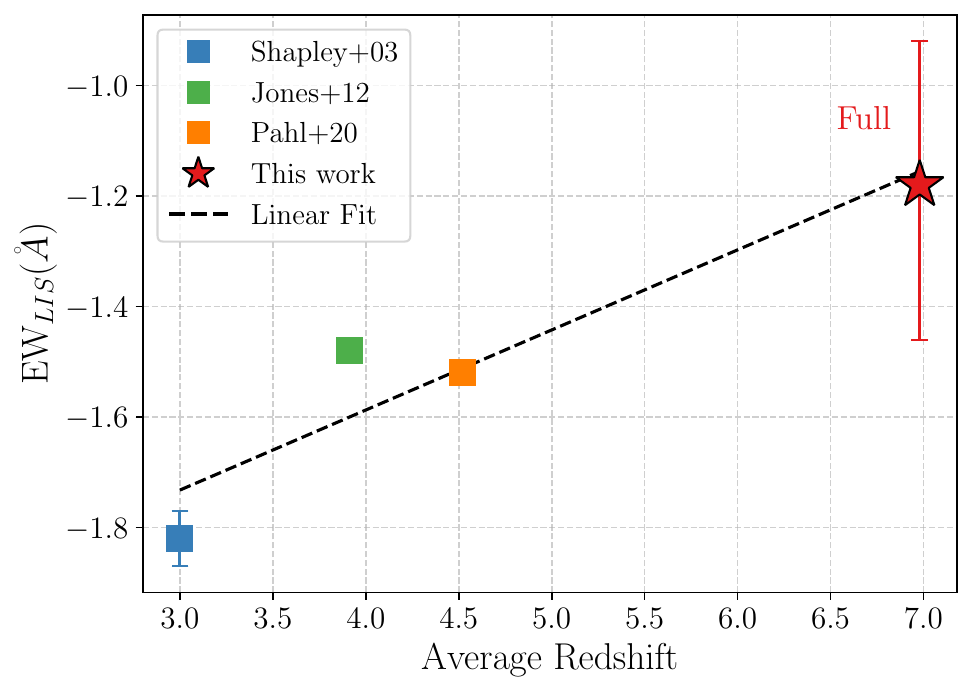}
\caption{The redshift trend of average $\rm{EW}_{\rm{LIS}}$ from $z=2$--5 samples (square markers) and our $z\sim7$ composite (star marker). $\rm{EW}_{\rm{LIS}}$ values are measured from skewed Gaussian profile fits applied consistently to all composite spectra. Plotted redshifts represent the average for each composite spectrum (or the reported median redshift in the case of \citealt{Pahl2020_z5stacks}). The black dashed line represents the best-fit linear relation, illustrating the overall trend of weaker $\rm{EW}_{\rm{LIS}}$ at higher redshifts.
}
\label{fig:EW_LIS_vs_redshift}
\end{figure}

The full PANCAKEZ stack shows weaker LIS absorption ($\rm{EW}_{\rm LIS} = -1.18^{+0.26}_{-0.28}$~\AA) compared to equivalent literature stacks at $z\lesssim4$ ($\rm{EW}_{\rm LIS} \lesssim -1.5 \AA$; square markers in Figure~\ref{fig:LIS_EW_vs_LYA_EW_UPDATED}), although our $z\sim7$ result is consistent with that at $z\sim4$ \citep{Jones2012_faintLBG} within the $1\sigma$ uncertainty. 
This may arise in part from the lower intrinsic UV  luminosities of our $z\sim7$ sample \citep[e.g.,][]{Du2018ApJ_specprops_LBGs}. Additionally, the weaker LIS absorption strength in the $z\sim7$ PANCAKEZ stack is indicative of a reduced average \fcovlis\ in these galaxies \citep[e.g.,][]{Jones2013_z4Lensed,Vasan2023}. Because \fcovlis\ correlates with \fcovhi\ \citep[e.g.,][]{Reddy2016ApJ_gascovfrac, Chisholm2018_fesc_predictions}, the PANCAKEZ galaxies could be exhibiting lower \fcovhi\ than the $z=2-5$ literature stacks. However, a caveat of the lower \fcovhi\ interpretation is that it assumes all HI gas in the ISM contains LIS metal ions to a similar degree. If there is pristine or very metal-poor \Hi\ gas along the line of sight, which is optically thin to the LIS transitions studied here, then this result should be interpreted as a lower limit on \fcovhi. We also note that \citet{Du2018ApJ_specprops_LBGs} found the average $\rm{EW}_{\rm LIS}$ generally decreases at higher redshift (i.e., less absorption) on average from $z\sim 2-4$. We plot $\rm{EW}_{\rm LIS}$ as a function of average redshift from respective composites in Figure~\ref{fig:EW_LIS_vs_redshift}, demonstrating that our result at $z\sim7$ is a continuation of this observed trend.

The PANCAKEZ measurements shown in Figure~\ref{fig:LIS_EW_vs_LYA_EW_UPDATED} are offset toward weaker LIS and/or weaker \Lya\ emission compared to the trend seen in $z=2$--5 samples. We attribute this difference to suppression of \Lya, in particular the decreased transmission of \Lya\ through the IGM at $z>6$. As discussed in \citet{Pahl2020_z5stacks}, the LIS and Ly$\alpha$ anti-correlation remains roughly constant from $2 \lesssim z \lesssim 4$. \citet{Pahl2020_z5stacks} shows that the observed relation deviates at $z\gtrsim4.5$ but can be recovered at higher redshifts when accounting for IGM attenuation. 
The PANCAKEZ measurements of EW(\Lya) shown in Figure~\ref{fig:LIS_EW_vs_LYA_EW_UPDATED} (solid red stars) are not corrected for IGM attenuation, which explains their offset. A recent study by \cite{Tang2024_Lya_SFGs_z613} quantified the IGM attenuation by comparing the distribution of EW(\Lya) for galaxies in the EoR with a reference sample at $z\sim5$. They report an average of $\sim 50\%$ reduction on {\lya} transmission between $z=6.5$--8 and $\sim 26\%$ at $z=8$--10. Applying this correction to the full PANCAKEZ stack (which predominantly falls within $z=6.5$--8), our measured EW(\Lya) doubles to $\sim 9 \; \mathrm{\AA}$. We plot this IGM-corrected PANCAKEZ result with a faded red star in Fig~\ref{fig:LIS_EW_vs_LYA_EW_UPDATED} along with an arrow illustrating the directional shift. The corrected PANCAKEZ stack still lies offset from the lower-$z$ trends, but the overall shift indicates that agreement can be achieved with a somewhat larger IGM correction. While we applied a $50\%$ IGM correction, a comparison between the $z\sim5$ and $z\sim7$ stacks suggests EW(\Lya) decreases by $\sim$70\%. Overall, while the comparisons between composite spectra and \cite{Tang2024_Lya_SFGs_z613} are based on different methodology, we can conclude that observed \Lya\ emission in the full PANCAKEZ composite spectrum is heavily attenuated by $\sim$50--70\% due to the IGM.

\subsubsection{Weak outflows at $z>6$}
Galactic-scale outflows are a common feature of SFGs with high surface density of star formation. At $z\gtrsim2$, SFGs show ubiquitously strong outflows \citep[e.g.,][]{Steidel1996_z3_SFGspec, Franx1997_gravlen_z4p9, Pettini2001_LBGs, Pettini2000_UVspec_LBG, Shapley2003_z3LBG, SaldanaLopez_2023_VANDELS_ISM} and observed trends from $z=2-5$ suggest a similar picture at higher $z$. Typically identified with blueshifted LIS absorption and/or redshifted {\lya} emission, velocity offsets of these spectral features give insight into the gas kinematics within the host galaxy. In this work, LIS and Ly$\alpha$ velocity offsets are quantified using velocity centroids ($v_{\rm{cen}}$) measured from respective line profile fits (discussed in Section~\ref{sec:LIS_measurement} and listed in Tables \ref{tab:LISMeasurements} and \ref{tab:LyaMeasurements}). 
Here $v_{\rm{cen}}$ represents the mean for both the normal and skewed Gaussian fits. For the skewed Gaussian case, LIS absorption $v_{\rm{cen}}$ is at smaller (more negative) velocities compared to the peak due to the negative skewness \citep[e.g.,][]{Vasan2023}. 

\begin{figure}
\centering
\includegraphics[width=\columnwidth]{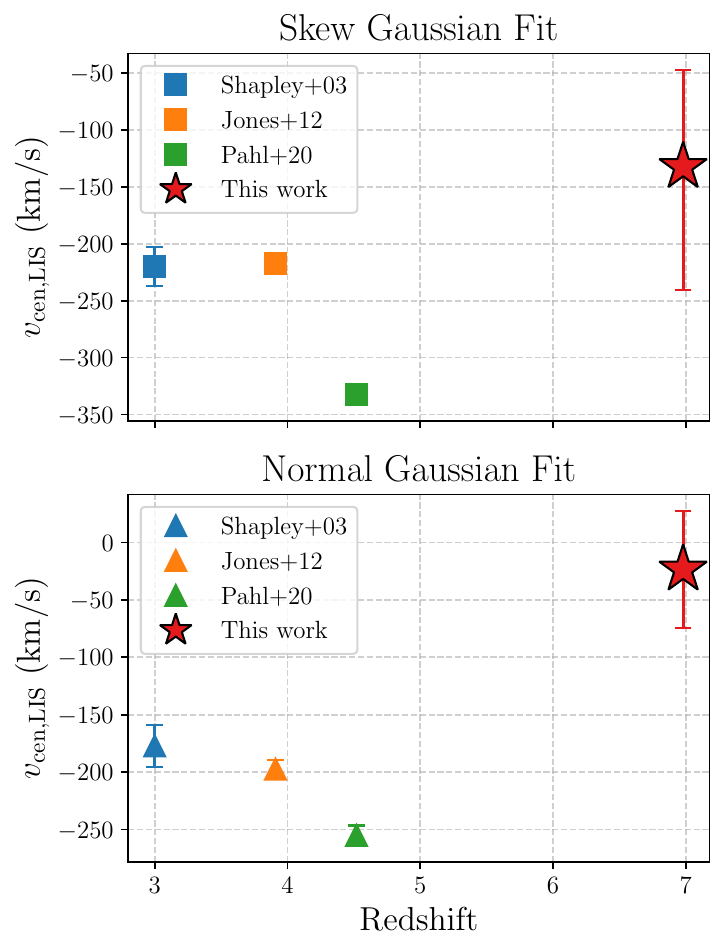}
\caption{The measured LIS absorption velocity centroid ($v_{\rm{cen,LIS}}$) for our full $z\sim7$ composite compared with $z=2$--5 samples. The top panel shows measured values from our best-fit skewed Gaussian models while the bottom panel shows the values measured from our best-fit symmetric models. Overall we see that the $z\sim7$ stack has a considerably lower LIS velocity offset compared to the lower-$z$ stacks. 
}
\label{fig:vLIS_vs_redshift}
\end{figure}

Figure~\ref{fig:vLIS_vs_redshift} shows our measured LIS velocity centroids from the $z\sim7$ stack as well as lower-redshift reference stacks \citep{Shapley2003_z3LBG, Jones2012_faintLBG, Pahl2020_z5stacks}. Comparing these studies directly, we see that $v_{\rm{cen,LIS}} \approx -200 \; \rm{kms}^{-1}$ across the lower-$z$ stacks ($z\sim 3$--5). Regardless of the fitted model (symmetric or skewed Gaussian), the PANCAKEZ stack shows a smaller offset (e.g., $v_{\rm{cen,LIS}} = -23\pm51$~\kms\ for the symmetric Gaussian fit). The smaller blueshift of ISM absorption implies that outflows at $z\sim7$ are weaker than those observed in $z\approx2$--6 SFGs. Furthermore, ISM absorption in the $z\sim7$ stack is consistent with no outflows ($v_{\rm{cen,LIS}} \approx 0~\kms$) at the $1\sigma$ level. However the LIS absorption FWHM is larger than for nebular emission lines (Tables~\ref{tab:LISMeasurements}, \ref{tab:nebEmissionLines}), suggesting outflowing and/or inflowing gas. The weak-outflow scenario could align with recent theoretical studies \citep{Dekel2023_FFBs, Li2024_FFB_obserpredicJWST} which discuss the possibility of high-$z$ galaxies with extremely high star formation efficiencies (also referred to as feedback-free starbursts). In these specific scenarios, star-forming clouds have free-fall times $\lesssim 1~\rm{Myr}$ due to high gas density ($\gtrsim 3\times 10^3 ~ \rm{cm}^{-3}$). Considering the $\sim 2~\rm{Myr}$ time delay between a starburst and the onset of effective feedback, such shortened free-fall times could allow star formation to proceed unhindered by stellar winds and supernovae, leaving little or no discernible outflow signature. 

In our analysis we have taken care to measure $v_{\rm{cen}}$ consistently for both PANCAKEZ and the lower-redshift composites (Tables~\ref{tab:LISMeasurements} and \ref{tab:LyaMeasurements}). The result of a lower average outflow velocity at $z\sim7$ is thus robust to measurement systematics. We note that our method fits the absorption with a single profile, as opposed to separate components for the ambient ISM (at $v=0$) and outflows/inflows. 
The $v=0$ ISM absorption component is expected to shift the overall profile towards lower velocities such that our measured $v_{\rm{cen,LIS}}$ values underestimate the true outflow velocity \citep[as discussed in, e.g.,][]{Steidel2010_cgmkinematics,Weldon2022_outflows}.

The modest outflow velocities seen in the PANCAKEZ stack are comparable to some galaxy samples studied at $z\sim2$ (with $v_{\rm{cen,LIS}} \approx 60$~\kms; e.g., \citealt{Weldon2022_outflows,Kehoe2024_outflows_z2}), which may be explained by increased absorption at the systemic velocity relative to galaxies at $z\gtrsim3$. However, increased systemic absorption does not explain the relatively small blueshifts in our $z\sim7$ sample given the weak total LIS absorption (i.e., low equivalent width). Instead we require decreased velocity and/or covering fraction of outflowing gas to explain the smaller blueshift at $z\sim7$. We thus attributed the evolution in LIS absorption centroids to weaker outflows in the PANCAKEZ stack relative to lower-$z$ galaxies.

\subsection{Intrinsic differences between LAEs and non-LAEs} 
\label{sec:discussion_LAEs_nonLAEs}

\begin{figure*}
\centering
\includegraphics[width=\textwidth]{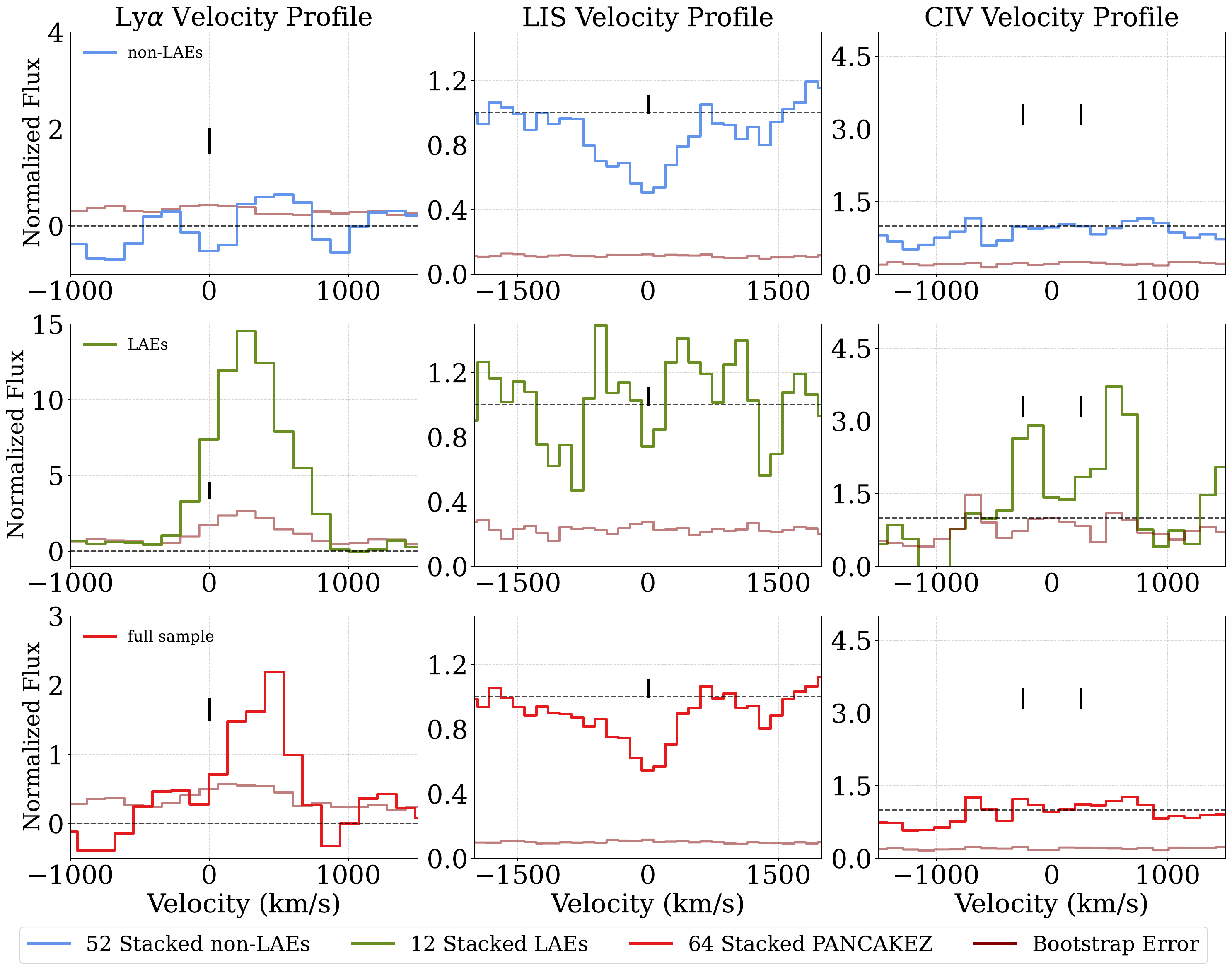}
\caption{Velocity profiles of Ly$\alpha$ emission (left column), LIS absorption (middle column), and nebular \Civ$\lambda\lambda$1548,1551 emission (right column) from our stacked spectra. Each row displays the respective profiles from our non-LAEs (top row), LAEs (middle row), and full PANCAKEZ (bottom row) samples. Bootstrap error spectra (maroon) are shown in each panel along with a dashed black line which represents the relative continuum level. Black tick marks denote the systemic redshift of the spectral features. Note that the {\Lya} profiles (left column) have different vertical scaling across the three stacks. This figure highlights key spectral differences seen between our full, LAE, and non-LAE stacks. In particular the LAE stack has weaker LIS absorption and stronger \Civ\ emission.
}
\label{fig:velocity_profiles_allstacks}
\end{figure*}

It is now well established that the fraction of star-forming galaxies exhibiting strong \Lya\ emission decreases in the EoR \citep[$z\gtrsim6$; e.g.,][]{Schenker2012,Schenker2014,Mason2019,Whitler2024,Tang2024_Lya_SFGs_z613}. We indeed see this effect directly in the relatively weak \Lya\ emission of the $z\sim7$ composite spectra presented here (Section~\ref{sec:LIS_Lya}). 
This effect is widely attributed to \Hi\ in the IGM. One expected consequence is that galaxies with \Lya\ emission are likely to reside in more highly ionized regions of the IGM, and can thus be used to identify reionized ``bubbles'' \citep[e.g.,][]{Tang2024_Lya_SFGs_z613,Mason_Gronke_2020_reionizedbubbles}. However, \Lya\ emission is also strongly regulated by intrinsic galactic and CGM properties (e.g., \citealt{Shapley2003_z3LBG}; Section~\ref{sec:LIS_Lya}). It is thus of interest to assess the extent to which observable \Lya\ emission in the reionization epoch is driven by local galaxy properties in addition to the IGM. We demonstrate in this section that the LAE and non-LAE galaxies in our sample show significant intrinsic differences which affect their emergent \Lya\ emission. 

\subsubsection{Higher $\xi_{ion}$ and harder ionizing spectra in LAEs}

Nebular emission features indicate differences in the ionizing sources in the LAE and non-LAE composite spectra. Our sample selection requires detection of nebular emission in order to measure systemic redshifts, but the selection is independent of line flux ratios or ionization properties.
As discussed in Section~\ref{sec:neblinediagonstic_O32_R23}, the LAE spectrum shows higher ionization (i.e., higher O32 index) and lower gas-phase metallicity than the non-LAEs. The emission equivalent widths are also larger, with EW(\Hb)~$= 168\pm4$~\AA\ for the LAE composite being approximately twice that of the non-LAEs ($74\pm1$~\AA). 
The rest-UV emission lines likewise show striking differences, in particular with strong nebular \Civ\ emission in the LAE composite (Figure~\ref{fig:velocity_profiles_allstacks}). 
\cite{Kumari2024_LAEstacks} similarly report a higher average EW(\Hb) and EW(\Civ) for $z>6$ LAEs, although their sample exhibits overall higher EW(\Hb) than in PANCAKEZ.
A double Gaussian fit of the  \Civ\ profile  gives an EW(\Civ)~$= 11\pm4$~\AA\ for the LAE stack (and $0.3\pm1.0$~\AA\ for the non-LAEs). We caution that \Civ\ line profiles may be affected by resonant scattering and modest signal-to-noise. Our fitted model reports redshifts of $60\pm30$~\kms\ for \Civ$\lambda$1548 and $280\pm20$~\kms\ for \Civ$\lambda$1551 relative to systemic, with a line ratio $\lambda$1551/$\lambda$1548~$=1.4\pm0.4$ (in contrast to the intrinsic ratio of 0.5). 
The flux ratio and different redshifts could arise from interstellar absorption and scattering, from stacking spectra with different emergent line profiles, and/or from noise fluctuations \citep{Berg2019}. The intrinsic equivalent width could plausibly range from $\sim$5~\AA\ based on the weaker emission peak which we attribute to \Civ$\lambda$1548 (e.g., if the apparent $\lambda$1551 feature is spurious), or as high as $\gtrsim$20~\AA\ based on $\lambda$1551 (e.g., if the $\lambda$1548 line has been attenuated by resonant scattering effects). 
Regardless, the LAE composite reveals nebular \Civ\ emission in significant contrast to the non-LAEs, with a redshift suggesting that \Civ\ is scattered by ionized outflowing gas. 

If the hypothesis that \Lya\ emission from galaxies in the EoR is regulated entirely by the surrounding IGM, we might expect little difference in the ionizing properties of LAE and non-LAE sources. In contrast we find that emergent \Lya\ emission from the LAE sample is directly related to ionizing radiation production within the galaxies themselves. The large EW(\Hb) and nebular \Civ\ indicate that the LAEs in our sample are dominated by very young and metal-poor stellar populations \citep[e.g.,][]{Senchyna2022}, relative to the non-LAEs. Moreover, the larger \Hb\ equivalent width is associated with higher ionizing photon production rates $\xi_{ion}$ (e.g., \citealt{Robertson2010_ionizationEQ,Chevallard2018}; see also \citealt{Pahl2020_z5stacks} regarding higher $\xi_{ion}$ in LAEs at $z\sim5$), while strong \Civ\ emission reveals a harder ionizing spectrum in the LAEs \citep[e.g.,][]{Kumari2024_LAEstacks}. 
The hard radiation produced by LAEs may indeed contribute to more efficiently ionizing the surrounding IGM, and thus enable \Lya\ to be transmitted \citep[e.g.,][]{Tang2024_Lya_SFGs_z613}. However, we emphasize the conclusion that observability of \Lya\ emission in our $z\sim7$ sample appears to be significantly linked to galactic properties in addition to the state of the surrounding IGM. This effect should be accounted for in any analysis of IGM neutral fraction tomography based on galaxies' \Lya\ emission. 

\subsubsection{Weaker LIS absorption in LAEs}

The equivalent width of \Lya\ emission is strongly correlated with LIS absorption \citep[e.g., Figure~\ref{fig:LIS_EW_vs_LYA_EW_UPDATED};][]{Shapley2003_z3LBG,Du2018ApJ_specprops_LBGs}. This arises naturally from \Hi\ and dust which are physically associated with the low-ionization metal absorption, leading to scattering and absorption of \Lya\ respectively. 
We find the same general correlation within our $z\sim7$ sample (Figure~\ref{fig:velocity_profiles_allstacks}; Table~\ref{tab:LISMeasurements}): the LAE composite spectrum exhibits weaker LIS absorption (EW$_{\mathrm{LIS}}=0.27\pm0.42$~\AA) than the non-LAE (EW$_{\mathrm{LIS}}=1.35\pm0.26$~\AA) and full composites. 
All of our composite spectra show evidence of suppressed \Lya\ emission (e.g., Figure~\ref{fig:LIS_EW_vs_LYA_EW_UPDATED}) which we attribute to IGM attenuation (Section~\ref{sec:LIS_Lya}). Nonetheless, the 2-$\sigma$ correlation of EW$_{\mathrm{LIS}}$ with \Lya\ emission suggests that galaxies' interstellar and circumgalactic media play a significant role in the observability of \Lya\ emission in the EoR, similarly to that of galaxies at $z<6$.

\subsection{ISM properties and implications for Lyman continuum escape in the Epoch of Reionization}
We now consider the broad physical characteristics of the diffuse ISM seen in absorption for our sample, and its redshift evolution into the EoR. The LIS absorption lines studied here are thought to predominantly arise from cool ($T\sim10^4$~K) and largely neutral (\Hi) gas in the ambient ISM and in galactic-scale outflows \citep[e.g.,][]{Tumlinson2017}. 
While the LIS and \Hi\ covering fractions are not identical, LIS absorption has indeed been shown to trace \Hi\ and \fesc\ in samples at lower redshifts \citep{Reddy2016ApJ_gascovfrac,Chisholm2018_fesc_predictions}.
Our stacking analysis extends earlier work at lower redshifts, providing equivalent measurements for the UV-luminous galaxy population over the range $2 \lesssim z \lesssim 7$ \citep[e.g.,][]{Du2018ApJ_specprops_LBGs,Pahl2020_z5stacks}. The average LIS absorption equivalent width decreases at higher redshifts, and we find a smooth continuation of this trend at $z\sim7$ (see Fig~\ref{fig:EW_LIS_vs_redshift}). We recover a correlation of weaker LIS absorption with stronger \Lya\ emission, as established at lower redshifts, but with overall weaker \Lya\ in our sample which is presumably due to neutral \Hi\ in the IGM at $z\simeq7$. 
While composite spectra at $z=3$--5 show LIS centroid velocities blueshifted by $\sim$200--300~\kms, our $z\sim7$ composite has a factor $\gtrsim$2 smaller blueshift (see Fig~\ref{fig:vLIS_vs_redshift}) -- a rapid change within $<$500 Myr compared to $z\sim5$. 
Smaller blueshifts could be explained by outflows which are more highly ionized (thus having weaker LIS absorption), more collimated (thus having lower covering fractions and weaker absorption), or lower bulk velocity. Overall the smaller blueshift indicates less efficient removal of cool interstellar gas. 

The weak LIS absorption lines in our composite spectra suggest a relatively patchy or porous \Hi\ distribution. 
At lower redshifts ($z<6$), galactic outflows are often invoked as a cause of patchy ISM. However, we observe modest outflow effects with a relatively small blueshift of interstellar absorption lines. An intriguing possibility is that the efficiency of star formation may be relatively high in the $z\sim7$ galaxies which comprise our sample, with conversion of gas into stars having a significant role in depleting the ISM \citep[e.g.,][]{Dekel2023_FFBs}. In such a scenario there could also be less cool gas entrained in the star formation-driven outflows. 
In any case, a patchy neutral ISM is more conducive to the escape of both \Lya\ and Lyman continuum photons. 
For post-reionization era galaxies, the LIS absorption equivalent width in our composite spectra is typically associated with strong \Lya\ emission. 
In fact we would expect the intrinsic \Lya\ emission to be stronger in our $z\sim7$ composite than in the lower-redshift comparison samples, and we attribute the weak observed \Lya\ to neutral \Hi\ in the IGM. 
Similarly the weak LIS absorption provides indirect evidence for substantial escape of Lyman continuum radiation from our sample. 
Indeed, composite spectra of $z\simeq3$ galaxies with comparable LIS absorption (EW(LIS)~$\approx 1.0$--1.4~\AA) have strong \Lya\ emission (EW~$\simeq20$--45~\AA) and absolute ionizing escape fractions $f_{esc} \simeq 0.15-0.3$ \citep{Steidel2018_KLCS_z3}. 
Our inference of moderately high $f_{esc}$ at $z\sim7$ is similar to independent analyses based on nebular emission properties of galaxies in the same redshift range \citep[e.g.,][]{Hu2024_ISM_stacks}. ISM absorption properties thus represent a complementary constraint on ionizing escape fractions, by probing the cool ISM which is directly responsible for regulating $f_{esc}$. 

\section{Summary} \label{sec:conclusion}
In this study, we present a comprehensive spectral stacking analysis of galaxies confirmed at $z_{\rm spec}\geq6$. Utilizing JWST/NIRSpec medium resolution data, we construct composite spectra from a sample of $64$ galaxies which are representative of moderately luminous SFGs in the epoch of reionization. 
We constructed three composite stacks from this sample: (i) a full stack of all $64$ galaxies, (ii) a LAE stack consisting of the subsample with detected \Lya\ emission ($N = 12$), and (iii) a non-LAE stack consisting of the subsample of non-LAEs ($N=52$). Our analysis is focused on rest-UV LIS absorption and transmitted \Lya\ emission, as well as rest-optical nebular emission lines in the composite spectra. We compare our results to lower-$z$ samples and discuss implications for EoR galaxies. We summarize our key findings below:

\begin{itemize}    
    \item LIS absorption at $z\sim7$ is relatively weak ($\rm{EW(LIS)} = -1.18 \pm 0.26 \AA$) compared to the lower-$z$ studies ($\rm{EW(LIS)} \lesssim -1.5 \AA$). We attribute this difference to lower LIS covering fraction {\fcov} in the $z\sim7$ PANCAKEZ sample. The lower LIS {\fcov}, which is directly connected to the \Hi\ {\fcov}, suggests a more porous ISM which allows for more ionizing photons to escape into the IGM. 
    A reduced \Hi\ {\fcov} and consequently increased ionizing escape fraction \fesc\ supports the hypothesis that SFGs are the main driver of cosmic reionization at $z>6$. 

    \item The full PANCAKEZ stack deviates from the anti-correlation trends between EW(LIS) and EW({\lya}) demonstrated by works at lower redshift. As expected, our $z\sim7$ galaxies residing in a partially neutral IGM exhibit suppressed Ly$\alpha$. While we do not model or correct the suppressed EW(Ly$\alpha$) in our stacks, we do find that our transmitted {\lya} at $z\sim7$ is $\approx 20\%$ the strength as measured in \citet{Pahl2020_z5stacks} at $z\sim5$.

    \item While the entirety of the PANCAKEZ sample is affected by IGM attenuation, LAE and non-LAE sub-samples show the physically linked behavior between LIS absorption and {\lya} emission. Our LAE stack reports EW(LIS)$\approx 0.27 \pm 0.4 \AA $ which implies low HI and dust presence explaining the elevated detection of Ly$\alpha$. Similarly the high EW(LIS)$ \approx1.4 \pm 0.3 \AA$ found in the non-LAE stack translates to higher HI reducing the overall transmitted Ly$\alpha$ via scattering and absorption. Even at EoR redshifts, we see the physical link between LIS absorption and {\lya} emission. 

    \item We find evidence for intrinsic differences between our LAEs and non-LAEs subsamples. The LAE stack shows extremely high EW(\Hb)~$= 168\pm4$~\AA\ which is more than double what is produced by the non-LAEs (EW(\Hb)~$= 74\pm1$~\AA). Additionally, nebular CIV emission is only found in the LAE stack (EW(\Civ)~$= 11\pm4$~\AA), with both the non-LAE stack and full PANCAKEZ stack showing no detection of the high ionization line. The high {\Hb} and detected CIV suggest that our LAEs have hard radiation fields and enhanced $\xi_{ion}$; which is consistent with the current picture of LAEs residing in local ionized bubbles. 
    
    The contrast between the LAEs and non-LAEs spectra is striking. In the absence of the IGM, we would expect LAEs and non-LAEs to be intrinsically different. When comparing the two subgroups in a partially neutral IGM, we could expect there to be some overlap because the IGM suppresses the transmission of Ly$\alpha$. The PANCAKEZ sub-samples, however, show drastic differences linking younger, metal-poor stars to our LAEs. We conclude that our Ly$\alpha$ emission appears to be considerably linked to galaxy properties which should be taken into account for any future efforts using LAEs to asses the IGM neutral fraction tomography. 

\end{itemize}

\section{Acknowledgments}
We acknowledge support from NASA grants JWST-GO-01914 and HST-GO-16667. We thank Dan Stark for valuable discussions. TJ acknowledges support from a UC Davis Chancellor's Fellowship. RSE acknowledges generous financial support from the Peter and Patricia Gruber Foundation.  MB acknowledges support from the ERC Grant FIRSTLIGHT, Slovenian national research agency ARIS through grants N1-0238 and P1-0188, and the program HST-GO-16667, provided through a grant from the STScI under NASA contract NAS5-26555.

\bibliography{PANCAKEZ_2024}{}
\bibliographystyle{aasjournal}

\end{document}